%% file: fpe_main.tex
\documentclass[12pt,a4j,notitlepage,fleqn]{article}

\usepackage[verbose]{geometry}
\usepackage[symbol]{footmisc}
\usepackage[T1]{fontenc}
\usepackage{authblk}

\makeatletter

\newcommand{\fboxsubsec}[1]{
	\begin{flushleft}
		#1
	\end{flushleft}
	}
\newcommand{\fboxsubsubsec}[1]{
	\begin{flushleft}
		#1
	\end{flushleft}
	}
\renewcommand{\subsection}{\@startsection{subsection}{2}{0pt}
	{1ex}
	{0.5ex}
	{\reset@font\it\fboxsubsec}
	}
\renewcommand{\subsubsection}{\@startsection{subsubsection}{2}{0pt}
	{1ex}
	{0.5ex}
	{\reset@font\fboxsubsubsec}
	}
\makeatother

\title{The Futures Premium and Rice Market Efficiency\\
in Prewar Japan}%

\author{Mikio Ito$^{a}$, \ Kiyotaka Maeda$^{b}$ \ and \ Akihiko Noda$^{c,d}$\thanks{\scriptsize Corresponding Author. E-mail: noda@cc.kyoto-su.ac.jp, Tel. +81-75-705-1510, Fax. +81-75-705-3227.}

{\scriptsize ${}^{a}$ \it Faculty of Economics, Keio University, 2-15-45 Mita, Minato-ku, Tokyo 108-8345, Japan}

{\scriptsize ${}^{b}$ \it Faculty of Economics, Seinan Gakuin University, 6-2-92 Nishijin, Sawara-ku Fukuoka 814-8511, Japan}

{\scriptsize ${}^{c}$ \it Faculty of Economics, Kyoto Sangyo University, Motoyama, Kamigamo, Kita-ku, Kyoto 603-8555, Japan}

{\scriptsize ${}^{d}$ \it Keio Economic Observatory, Keio University, 2-15-45 Mita, Minato-ku, Tokyo 108-8345, Japan}}
\date{Forthcoming in {\it Economic History Review} (DOI:10.1111/ehr.12608) \empty}

\renewcommand\thefootnote{\arabic{footnote}}

\usepackage{graphicx} 

\usepackage[]{natbib}%
\usepackage{amsmath,amssymb}%
\usepackage{ascmac}%
\usepackage{multirow}%
\usepackage{lscape}%
\usepackage{subfigmat}

\usepackage{pifont}%
\usepackage{arydshln}%
\usepackage[format=hang]{caption}
\usepackage[all]{xy}
\usepackage{url}
\bibpunct{(}{)}{;}{a}{}{,}

\def\hsymbu#1{\smash{\lower1.7ex\hbox{\huge$#1$}}}

\def\ve #1{{\mbox{\boldmath $#1$}}}

\newcommand{\citetapos}[1]{\citeauthor{#1}'s \citeyearpar{#1}}

\def\ve #1{{\mbox{\boldmath $#1$}}}

\begin{document}

\begin{titlepage}

\renewcommand{\thepage}{}
\renewcommand{\thefootnote}{\fnsymbol{footnote}}

\maketitle

\vspace{-10mm}

\noindent
 \hrulefill

\noindent
{\bfseries Abstract:} This paper studies the interrelation between spot and futures prices in the two major rice markets in prewar Japan from the perspective of market efficiency. Applying a non-Bayesian time-varying model approach to the fundamental equation for spot returns and the futures premium, we detect when efficiency reductions in the two major rice markets occurred. We also examine how government interventions affected the rice markets in Japan, which colonized Taiwan and Korea before World War I\hspace{-.1em}I, and argue that the function of rice futures markets crucially depended on the differences in rice spot market's structure. The increased volume of imported rice of a different variety from domestic rice first disrupted the rice futures. Then, government intervention in the rice futures markets failed to improve the disruption. Changes in colonial rice cropping successfully improved the disruption, and colonial rice was promoted to unify the different varieties of inland and colonial rice.\\

\noindent
{\bfseries Keywords:} Rice Futures Markets; Futures Premium; Market Efficiency; Non-Bayesian Time-varying Model Approach.\\

\noindent
{\bfseries JEL Classification Numbers:} N25; G13; C22.

\noindent
\hrulefill

\end{titlepage}

\bibliographystyle{asa}%

\pagebreak

\input{fpe_intro}

\input{fpe_history}

\input{fpe_model}

\input{fpe_data}

\input{fpe_empirical}

\input{fpe_interpretation}

\input{fpe_conclusion}

\input{fpe_ack}

\clearpage

\input{fpe_main.bbl}

\input{fpe_table}

\end{document}

%% file: fpe_intro.tex
\section{Introduction}\label{fpe_sec1}

Price risks associated with lean crop yields, wars, and economic crises are persistent problems. Historically, global futures markets have been established to hedge such risks. In general, a futures market is characterized as a well-organized forward market of a commodity, and it provides a fine index of the expected commodity price. However, it is not clear how and to what extent government interventions affected both the futures and physical markets of agricultural commodities of colonized countries before World War I\hspace{-.1em}I.

Steam locomotives and steamships became popular after the Industrial Revolution. These new transportation channels provided mass, high-speed transportation. They also enabled rapid and expanding commercial transactions including foreign trade. However, the expanding trade with more distant locations exposed traders to greater price risks. Accordingly, futures markets were established in the 19th century, for example, in Chicago (founded in 1848), Frankfurt (founded in 1867), New York (founded in 1870), and London (founded in 1877) (see \citet[p.11]{kaufmann1984hfm}). The futures market in Japan was established in the early 18th century.

The Tokugawa Shogunate, which was formed in 1603, enjoyed a long period of rule throughout Japan and stabilized the internal economy and civic order. The stability of the 17th century allowed authorities to improve transportation infrastructure such as roads, coastal routes and harbors. The infrastructure improvements increased trade among remote locations. The Shogunate also permitted shipbuilders to build large merchant vessels under a stable military in 1638. This permission enabled mass transportation. Thus, Japan experienced expanding trade among remote locations beginning in the 17th century. While rice trading became more active, rice merchants faced greater price risk because of changing commercial transactions. Accordingly, the Shogunate certified the {\it{Dojima Kome Kaisho}} (i.e., the Osaka-Dojima rice exchange) in 1730.

The Osaka-Dojima rice futures market only traded domestic rice in the Tokugawa era. This limited trading was the result of severe restrictions on foreign trade and international relationships, called {\it{Sakoku}}, enforced by the Shogunate. The isolation policy began in the mid-17th century. However, the market began to trade domestic rice and imported rice after the late 19th century. We consider the consequences of the difference between the two eras with respect to rice futures.

Japan concluded the Treaty of Amity and Commerce with France, Netherlands, Russia, the United States, and the United Kingdom in 1858 and opened the ports in 1859. The Meiji government declared a new regime in 1868. Discarding the isolation policy of the previous authority, Japan adopted Western technologies and modernized transportation infrastructure for steam locomotives and steamships. The resumption of foreign trade and the introduction of modern infrastructure resulted in an increase in commercial transactions in Japan. Commodity futures trading also became more active. Rice was a major commodity in the futures markets and was traded in the two major commodity futures markets in Tokyo and Osaka. \citet{taketoshi1999ear} examines the efficiency of the Tokyo rice futures market using the \citetapos{fama1987cfp} model. \citet{shizume2011fep} studies the rational expectation formation of rice futures in the Osaka rice futures market using \citetapos{hamilton1987mfg} model. \citet{nakanishi2002cmm} and \citet{koiwa2003rmm} discuss the evolution of the two major rice futures markets considering the correlation of price and shares of trade changes over time. However, the previous literature has paid little attention to the characteristics of the rice market in prewar Japan and the nation's dependency on rice from its colonies.

In prewar Japan, the rice policy had great significance for food supply policy but also price stabilization policy. Rice has been the staple food since ancient times in Japan, and fluctuation in rice prices was the primary cause of fluctuation in general price levels. In fact, 13\% of the weight of the Tokyo Wholesale Price Index in the 1933 Base was the rice price. The weight of cotton yarn, the second largest weight, was only half that of the weight of rice (see \citet[p.40]{boj1987hys}). Accordingly, the Japanese government paid particular attention to the fluctuation in rice prices. For example, in 1890, Masayoshi Matsukata, the finance minister, mentioned, ``The rice price increases significantly more than the price of any other commodities, and people live in dire poverty. Alongside increases in rice prices, workers will force manufacturers to raise their wages. In addition, they may provoke riots or commit theft.''\footnote{See \citet[p.182]{ota1938rpp}.} That is, the government feared that a rising rice price would result in cost-push inflation and aggravate public security. Accordingly, rice imports grew after the 1890s in Japan. However, increasing rice imports caused an increase in the outflow of specie money. Consequently, the government imported rice from the Japanese colonies.

When Japan won the Japanese-Sino War, from the year 1894 to the year 1895, the country gained Taiwan as the first Japanese colony. In 1910, Japan annexed the Korean peninsula. In prewar Asia, Japan was the only country possessing colonies, and it imported food and resources from its colonies: rice, sugar, soybean, and salt. Japan was growing colonial trading. Since the 17th century, European countries owned colonies in the Americas, Africa, and Asia. The colonies supplied foods and resources to the home countries. However, Japan depended on its colonial trading to a greater extent than European countries (see \citet{okubo2007tbf} and \citet[pp.72--75]{hori2009hce}). In particular, domestic rice production had not met increasing demand from domestic consumers because of rapid industrialization and urbanization since the 1890s. Japan compensated for the insufficiency in domestic rice supply by importing colonial rice. However, few studies focus on the effect of colonial trading on commodity markets in prewar Japan.

\citet{ito2016meg} show that the efficiency of each of the Tokyo and Osaka rice futures markets varied over time using a non-Bayesian time-varying VAR model. The authors argue that government intervention allowing imported rice deliverables in the futures markets reduced efficiency. In prewar Japan, the consumer recognized that the quality of imported rice, including Taiwanese and Korean rice, was lower than domestic rice. In fact, \citet[pp.32-33]{sugihara1996tbm} notes that there was an obvious difference in price between domestic rice and imported rice, and the price gap increased until the early 1910s. In addition, \citet[pp.135--145]{mochida1970drm} mentions that Korean rice was mainly concentrated in Osaka. Hence, rice circulating in Osaka was different from the rice circulating in Tokyo. Although the previous studies elucidate that there was a difference in the rice spot market between Tokyo and Osaka, \citet{ito2016meg} ignore rice trading in the spot market because they use only the futures price data. Therefore, the authors do not discuss the interrelation between the two major rice markets for spot and futures rice prices.

This paper studies the interrelation between spot and futures prices in the two major rice markets in prewar Japan from the perspective of market efficiency. We then examine how government intervention disrupted and ameliorated the efficiency measures of the spot and futures rice markets. Our examination has four components. First, we estimate the well-known equation for spot returns and futures premiums and test its parameters to examine whether the markets were efficient. Second, we test the parameter constancy in the estimation procedure. Third, if there is doubt, we estimate the parameters of the equation at each period assuming that the two major rice markets are confronted with continuous structural change. At this stage, we use a non-Bayesian time-varying model approach, which allows us to estimate the parameters and to conduct statistical inference from a residual-based bootstrap technique (see \citet{ito2014ism,ito2016eme}). Fourth, we detect when the efficiency measure of the rice markets dropped and investigate the possibility that government interventions in the rice markets affected the efficiency measure. Then, we historically investigate how government interventions disrupted and ameliorated the efficiency measure of the rice markets.

This paper is organized as follows. Section \ref{fpe_sec2} provides a short historical review of rice markets in Japan. Section \ref{fpe_sec3} presents our methodology for our non-Bayesian technique for a linear regression equation with time-varying parameters. This section also provides statistical inference for the above time-varying parameters based on a residual-based bootstrap method. Section \ref{fpe_sec4} describes the data on the rice futures markets in prewar Japan. Section \ref{fpe_sec5} summarizes our empirical results of econometric analyses of time-varying regression model with the rice futures data. Section \ref{fpe_sec6} argues the background of the results and shows their historical interpretation. Section \ref{fpe_sec7} concludes.

%% file: fpe_history.tex
\section{A Historical Review of the Rice Markets in Japan}\label{fpe_sec2}

The Meiji government declared a new regime in 1868. The following year, the Meiji government forbade the Osaka rice futures market to trade because it considered the activity to be gambling. Rice traders requested the government to reopen the futures market. The government permitted the rice futures market in Osaka to resume trading in 1871 (see \citet[pp.26--40]{tsugawa1990sod}). In the same period, rice traders in Tokyo sought to establish another rice futures market, and two rice futures markets were set up in Tokyo. First, Hachirouemon Mitsui, a wealthy merchant in Tokyo, began futures rice trading through his company, {\it Boueki Shosha}, in 1871. {\it Chugai Shogyo Kaisha}, supported by representative merchants in Kagoshima, began rice futures trading in 1874. These two rice futures markets in Tokyo were unified in 1883 and were renamed Tokyo {\it Kome-Shokaisho} (i.e., the Tokyo Rice Exchange).\footnote{See \citet[pp.27--36]{tge2003htg} for details.} The two major exchanges in Tokyo and Osaka led the rice futures markets before World War I\hspace{-.1em}I in Japan from the 1880s.

Until the mid-1920s, the exchanges had not dealt with physical rice. The two major exchanges in Tokyo and Osaka began to allow rice traders to trade physical rice within the exchanges in Tokyo in 1923 and Osaka in 1928 (see \citet[pp.1148, 1156]{ota1938rpp}. Therefore, rice traders in the exchanges only traded in futures. There were three transactions with respect to different contract months in the Japanese rice futures markets; a nearby contract (one month), a second nearest contract (two months), and a deferred contract (three months). That is, the longest contract was three months. It was shorter than any other market worldwide; 12 months was the longest contract period in many futures markets. In fact, the longest contract period of representative futures markets, such as the wheat futures market in the Minneapolis Grain Exchange, the cotton futures market in the New York Cotton Exchange, and the Egyptian cotton futures market in the Liverpool Cotton Exchange was 12 months in the 1920s (see \citet[pp.67--68]{kamibayashi1935cm}). However, the Japanese government was concerned that the markets would treat rice futures with a longer contract period because futures trading had amounted to a gamble in the Osaka rice futures market in 1869 (see \citet[pp.67--97]{haneji1989sce}). In short, the basic framework of rice futures trading did not change in prewar Japan. In contrast, the circumstances around the rice spot markets changed frequently.

In the 1880s, rice exports to Europe increased. From 1884 to 1889, the amount increased from 0.5 million {\it koku} to 1.4 million {\it koku}.\footnote{{\it Koku} is a unit root of rice trading volume in Japan. One koku is equal to 180.39 liters.} Particularly, the main destination of rice exports was the United Kingdom. In fact, 50 per cent of rice exports from Japan was shipped to the United Kingdom in 1889 (see \citet[p.21]{omameuda1993fpm}). This situation arose from two causes. First, the yen against the pound increased in the 1880s. From the early to the late 1880s, the pound-yen exchange rate depreciated from four yen to seven yen (see \citet[p.34]{sugihara1996tbm}). Second, demand expanded while the supply did not increase in the European food market. In the late 1880s, the growing rice exports from Burma, the center of the rice supply to Europe, was stagnant. In addition, barley, wheat, and potato crops failed in France, Germany, and the United Kingdom in 1888 (see \citet[pp.29--30]{omameuda1993fpm}). Under the circumstances, the Japanese government directly exported rice.

In 1880, the government enacted {\it Biko Chochiku Ho} (the Emergency Rice Stocks Law) to stock rice in anticipation of a crop failure. This law allowed the government to buy and sell  rice physically in the spot market. Because the shelf life of rice was only a few years, the government had to sell some quantity of rice even if the rice crop in Japan was plentiful. However, the supply of rice exceeded demand in the late 1880s. From the early to the late 1880s, the volume of rice production increased by 15 per cent while the population grew by 6\% in Japan (see \citet[p.634]{toyo1927jss} and \citet[pp.1--2]{maf1909rcj}). The government did not sell the inventory of rice in the domestic market because it was concerned with the falling price of rice. Instead, the government sold rice in the foreign market to prevent the rice price from declining in the domestic market and to earn specie money. Specifically, the government exported rice actively from 1886 to 1889 (see \citet[p.165]{ota1938rpp}). During this period, the ratio of rice exported by the government to the total amount of rice export from Japan was 26\% (see \citet[pp.20--21]{omameuda1993fpm}). However, the terms of overseas rice trading changed again in the 1890s.

Since the 1890s, rice demand in Japan had exceeded its supply. Japan suffered from a chronic shortage of rice, and rice imports exceeded exports after the 1890s. In the 1890s, Japan imported rice from the Southeast Asian countries and Korea. However, beginning in the 1900s, rice imports from the Japanese colonies increased. Imported rice from the colonies was of a different variety than domestic rice. The former was an indica variety, and the latter was japonica, and each variety has a different texture and taste. When rice was shipped from Korea, farmers and brokers mixed sand, stones, and other impurities into rice sacks to cheat their trading partners by commodity weight. The rice traders dealing in Korean rice had to polish the rice with sand and stones. To this end, rice traders prepared dedicated milling machines that could decontaminate the sand and stones from the Korean rice. Particularly, the rice traders who had the milling machines for the Korean rice clustered in Osaka (see \citet[pp.591--598]{hishimoto1938skr}). Therefore, the rice from Korea was concentrated in Osaka because Korea is situated nearer Osaka than to Tokyo (see \citet[p.2]{maf1938rtf}). Finally, Korean rice trading in Japan at that time was characterized by, first, an imbalance in the regional distribution of Korean rice in Japan and, second, an increase in rice imports that diversified the distribution of rice in Japan.

Unlike the actual rice distribution, the rice futures markets continued to deal in domestic rice. However, the Ministry of Agriculture and Commerce first forced the exchanges to accept imported rice as an alternative to deliverable rice within 1890. This intervention and other similar interventions were aimed at suppressing the futures rice price when Japan faced a serious shortage of rice. Then, the futures markets often permitted rice traders to deliver imported rice of a different quality than domestic rice.

After the Russo-Japanese War (1904--1905), the rice price continued to increase although Japan addressed the rice shortage aggressively. The government attempted to suppress rice price increases and forced the futures markets to accept imported rice as a deliverable regularly in 1912. During the same period, the Governor-General of Korea promoted japonica rice cropping in Korea. This policy had gradually reduced quality differences between domestic and Korean rice since the late 1910s.

The nationwide riots, {\it kome-soudo} (the rice riots), occurred in 1918, spurred by a serious rice shortage caused by speculative holding-off by merchants who anticipated an increase in rice prices because of planned intervention by the Japanese Imperial Navy during the Soviet revolution and a self-fulfilling rise in price from this expectation. Reflecting on the disruption from the riot, the government began to intervene in the rice spot market and, in 1921, established the Rice Law. The first article of the Rice Law stated that if the government recognized the necessity to adjust rice supply and demand, it could buy, sell, change, process, and store the rice (see \citet[p.332]{ota1938rpp}). In short, the Rice Law allowed the government to buy and sell physical rice directly to adjust rice supply and demand. However, the government always intervened in the rice spot market by selling and buying only after estimating the supply and demand of rice (see \citet[pp.198--201]{omameuda1993fpm}). In 1925, the government amended the Rice Law to overcome the problems with supply and demand. Practically, the term ``rice supply and demand'', in the first article was replaced by ``the volume in circulation and price of rice.''\footnote{See \citet[pp.335--336]{ota1938rpp} for details.} That is, the government could flexibly respond to changes in price in the rice spot market. In the late 1920s, the Ministry of Agriculture and Forestry owned the physical rice in stock to reduce excessive volatility of rice price following the new Rice Law.\footnote{The Ministry of Agriculture and Commerce was separated into the Ministry of Agriculture and Forestry and the Ministry of Commerce and Industry in April 1925.}

%% file: fpe_model.tex
\section{The Model}\label{fpe_sec3}

This section presents our econometric method used to examine the time-varying structure of the Japanese rice futures market in the Meiji, Taisho, and early Showa periods (1868--1945). We employ a time- varying econometric model rather than testing statistically the efficient market hypothesis given the whole sample data to identify how and when the two rice futures markets changed in prewar Japan. In other words, the ordinary econometric models examine economic theories with historical data while our methods use a time-varying VAR model to explore historical events and the background that would change economic circumstances. Particularly, we examine when the rice futures markets faced structural changes in prewar Japan based on time-varying estimates of the VAR model.

The method is partly based on \citet{ito2014ism}, who analyze the dynamic linkages of stock prices among G7 countries and the efficiency of international stock markets. However, our method focuses on the dynamic relation between the rice futures premium and the returns on spot rice in prewar Japan; we examine the time-varying estimates of the fundamental equation for the futures premium. Employing a non-Bayesian time-varying model approach, we conduct statistical inference for the time-varying estimates using a residual-based bootstrap technique.

\subsection{Preliminaries}\label{subsec: prelim}
This paper considers the relation between the rice futures premium and the returns on spot rice. Following earlier studies (e.g., \citet{ito1993erf} and \citet{wakita2001edr}), we adopt the fundamental equation for the futures premium as our starting point:
\begin{equation}
(\log{S_{t+k}}-\log{S_t})=\alpha+\beta(\log{F_{t+k|t}}-\log{S_t})+u_t, \ \ (t=1,2,\cdots,T-k),
\label{eqn:fpe eq}
\end{equation}
where $S_t$ is the spot price at time $t$, $F_{t+k|t}$ is the futures price at time $t$ for delivery at time $t+k$, and $u_t$ follow an independent and identically distributed process. That is, we consider the $k$-th period forward contract. This type of regression equation is typically employed to test the unbiasedness hypothesis in the context of financial market efficiency. In practice, the null hypothesis is that $(\alpha,\beta)=(0,1)$ and $\{u_t\}$ are serially uncorrelated. Note that the hypothesis refers to the joint hypothesis of risk neutrality (or no-risk premium) and rationality.  In other words, the hypothesis states that no informed speculators can expect to make excess returns (see \citet{brenner1995act} for details).

In an efficient market, nobody accrues profit because of no arbitrage. In such a market, the coefficients $\beta$ in the fundamental equation (\ref{eqn:fpe eq}) is one, and $\alpha$ is zero. That is, there exists no systematic capital gain from futures trade, because the expected prediction error, $E[(S_{t+1}-S_t)-(F_{t+k|t}-S_t)]=E[S_{t+1}-F_{t+k|t}]$, is zero. On the other hand, in an inefficient market, some traders gain profits because of the existence of prediction error, which represents some arbitrage. In such a situation, the greater its deviation from one, the more some traders accrue profit from the arbitrage. Thus, we can consider the absolute deviation of $\beta$ from one as an efficiency measure of the market, and it allows us to detect inefficiency.


\subsection{Non-Bayesian Time-varying Regression Models}\label{subsec:state_space}
Suppose that there is concern for the time-varying structure of a possibly efficient futures market of the $k$-th period forward contract. The time series of spot returns $x_t=\log S_{t+k}-\log S_t$ and futures premiums $y_t=\log F_{t+k|t}-\log S_t$ should be examined. Note that $x_t$ and $y_t$ at $t$ are theoretically considered an unknown futures return and its predictor. In practice, attention is given to the differences between the two variables $x_t-y_t$ at each period to study the time-varying predictive performance of futures. However, this approach does not make sense. Because $x_t-y_t=\log S_{t+k}-\log F_{t+k|t},$ time series of the differences would exhibit a nearly white noise process with small values at a high probability. The reason is that $S_t$ and $F_{t+k|t}$ typically have a unit root, are highly correlated, and present spurious correlation. The above approach provides us with no information on the futures market.

We adopt another approach using \citetapos{ito2014ism} non-Bayesian time-varying model for Equation (\ref{eqn:fpe eq}). This approach estimates the coefficients in Equation (\ref{eqn:fpe eq}), possibly varying over time. We use the following equation:
\begin{equation}
 (\log{S_{t+k}}-\log{S_t})=\alpha+\beta_t(\log{F_{t+k|t}}-\log{S_t})+u_t, \ \ (t=1,2,\cdots,T-k).
\label{eqn:tv-fpe eq}
\end{equation}
This is a linear regression of the returns on spot rice on the rice futures premium. We consider whether $\beta$ varies with time; we regard $\alpha$ as time-invariant considering its insignificance in the preceding works. Equation (\ref{eqn:tv-fpe eq}) cannot be estimated because it is unidentifiable. Thus, many preceding works estimating a model with time-varying parameters assume a dynamic equation for such parameters, for example, random walk. We assume that $\beta_t$ follows a random walk represented as follows:
\begin{equation}
 \beta_{t+1}=\beta_t+v_t, \ \ (t=1,2,\cdots,T-k),
\label{eqn:rw_beta}
\end{equation}
where each $v_t$ follows an independent and identical distribution. We consider Equations (\ref{eqn:tv-fpe eq}) and (\ref{eqn:rw_beta}) together as a state space model. Following \citetapos{ito2014ism}, we employ a non-Bayesian technique to estimate the parameters in the state space model using time series of $S_t$ and $F_{t+k|t}$ as data.\footnote{See \citetapos{ito2014ism} Online Technical Appendix A.1, which is available at \url{http://at-noda.com/appendix/inter_market_appendix.pdf} for details.} We consider Equations (\ref{eqn:tv-fpe eq}) and (\ref{eqn:rw_beta}) together to estimate the parameters $\alpha,\beta_1,\cdots,\beta_{T-k}$ by using the following matrix form:
\begin{equation}
\left(%
\begin{array}{c}
 y_1\\ y_2\\ \vdots\\ y_{T-k}\\ \beta_0\\ 0\\ \vdots\\ 0
\end{array}
\right)
=\left(%
\begin{array}{ccccc}
1 &x_1 & & & \ve{O}\\
1 & &x_2 & &   \\
\vdots & & &\ddots &   \\
1 &\ve{O} & & & x_{T-k}\\
0 &1 &0 & & \ve{O}\\
0 &-1 &1 & & \\
\vdots & &\ddots &\ddots & \\
0 &\ve{O} & &-1 &1 \\
\end{array}
\right)
\left(%
\begin{array}{c}
 \alpha\\ \beta_1\\ \beta_2\\ \vdots\\ \beta_{T-k}
\end{array}
\right)
+
\left(%
\begin{array}{c}
 u_1\\ u_2\\ \vdots\\ u_{T-k}\\ v_1\\ v_2\\ \vdots\\ v_{T-k}
\end{array}
\right),
\label{eqn:matrix regression}
\end{equation}
where $x_t=\log{F_{t+k|t}}-\log{S_t}$, and $y_t=\log{S_{t+k}}-\log{S_t}$ for $t=1,2,\cdots,T-k$. This helps us to estimate the coefficients and to conduct statistical inference using a bootstrap technique shown in the next subsection. We estimate the state vectors $(\alpha \ {\beta}_1 \ {\beta}_2 \cdots {\beta}_{T-k})'$ at one time  based on  observations $x_1, x_2 \cdots, x_{T-k}, y_1, y_2 \cdots, y_{T-k}$ and a prior $\beta_0$ given by OLS or GLS that are familiar with economists. This vector estimate is identical to the Kalman smoother with a fixed interval of the corresponding state space model, Equations (\ref{eqn:tv-fpe eq}) and (\ref{eqn:rw_beta}).\footnote{See \citetapos{ito2016eme} Online Technical Appendix A.1, which is available at \url{http://at-noda.com/appendix/evolution_appendix.pdf} and \citet[ch.15]{maddala1998urc} for details.}\footnote{Our method for estimating random parameter regression models is different from the method in the literature of the 1970s (see \citet{swamy1970eir}, \citet{swamy1975bnb} and \citet{swamy1980lpe} for example). The latter study focuses on stationary stochastic coefficients such as an autoregressive moving average model (ARMA) process, whereas our study focuses on coefficients following a random walk, a non-stationary process.}

\subsection{Statistical Inference for Time-varying Parameters}\label{subsec:bootstrap}
This subsection presents our methodology for statistical inference on the time-varying estimates of $\beta_t$'s in Equation (\ref{eqn:tv-fpe eq}). The idea is so simple that we examine the {\it estimates} with the joint distribution of our time-varying {\it estimator} $\hat\beta_1,\hat\beta_2,\cdots,\hat\beta_{T-k}$ under the unbiasedness hypothesis that $\beta_t=1$ for any $t$ and $\alpha=0$. After setting a significance level, for example, 95\%, we construct upper and lower bounds based on the distribution as two time series: ${\beta^U_1,\beta^U_2,\cdots,\beta^U_{T-k}}$ and ${\beta^L_1,\beta^L_2,\cdots,\beta^L_{T-k}}$. Then, using a confidence band constructed from the two series of bounds, we examine in what period the $\hat{\beta}_t$ estimates are inside the band to identify the efficient market periods.

However, we cannot employ an asymptotic theory to construct the confidence bands on time-varying coefficients that are supposed to follow a random walk. This difficulty stems from the asymptotes involved in a Brownian motion or Brownian bridge, and it is complex to derive them theoretically; their distributions can only be represented with some Brownian motion or Brownian bridge if it is successful. That is, such distributions require a Monte Carlo technique to conduct some statistical inference. We are obliged to adopt a residual-based bootstrap technique to construct parameter confidence bands of a state space model under the assumption of the unbiasedness hypothesis: $(\alpha,\beta_1,\cdots,\beta_{T-k})=(0,1,\cdots,1)$ for our time-varying model. We consider the case where each $\beta_t$ is estimated by the above method while the data are generated by Equation (\ref{eqn:fpe eq}) when $(\alpha,\beta)=(0,1)$.

In practice, our residual-based bootstrap technique consists of the following steps.\footnote{See \citetapos{ito2014ism} Online Technical Appendix A.3, for which the basic idea is found in \citetapos{lutkepohl2005nim} Appendix D.3.} First, we fit Equation (\ref{eqn:fpe eq}) to the observed data. Then, we obtain a residual sequence $D_0=(\hat u_1,\hat u_2,\cdots,\hat u_{T-k})$. Second, we extract $N$ bootstrap samples $D_i=(\hat{u}^{(i)}_1,\cdots,\hat{u}^{(i)}_{T-k}),i=1,2,\cdots,N$ with a replacement from $D_0$ considering it an empirical distribution of the residuals. Third, we generate $N$ sets of $(\hat{u}^{(i)}_1,\cdots,\hat u^{(i)}_{T-k})$ for $i=1,2,\cdots,N$ using $D_i$ for $i=1,2,\cdots,N$. Fourth, we estimate $N$ sets of the time-varying coefficients and their corresponding residuals by applying our time-varying model (\ref{eqn:matrix regression}) to $N$ bootstrap data $(x_1,x_2,\cdots,x_{T-k},y^{(i)}_1,y^{(i)}_2,\cdots,y^{(i)}_{T-k})$ and a prior $\beta_0$ given for $i=1,2,\cdots,N$. Finally, we construct a confidence band of $(\alpha,\beta_1,\cdots,\beta_{T-k})$.

%% file: fpe_data.tex
\section{Data}\label{fpe_sec4}
We utilize the weighted average monthly data on the rice spot and futures prices in prewar Japan. For the rice futures data, three contract months of the Tokyo and Osaka rice exchanges are available: a nearby contract (one month), a second nearest contract (two months), and a deferred contract (three months). For the rice spot data, spot prices in Tokyo are on the Tokyo-Fukagawa rice spot market, and the spot prices in Osaka are the wholesale prices of rice in Osaka. These datasets consist of the following three statistics: (i) \citet{nakazawa1933nbh} for all futures prices from October 1880 to November 1932, (ii) the {\it Tokyo City Statistics} and \citet[p.342]{mci1931swp} for the spot prices in Tokyo from April 1881 to November 1932, and (iii) \citet{miyamoto1979owp} and the {\it Osaka City Statistics} for the spot prices in Osaka from April 1881 to November 1932.\footnote{Note that the rice futures data of nearby and nearest contract months in Tokyo have different starting dates from the rice futures data of the deferred contract month. Particularly, the data for the nearby contract month are available from January 1888, and the data for the nearest contract month are available from January 1898.} There are a few missing values in both sets of statistics. Therefore, we fill in the missing values using a seasonal Kalman filter. We take the first difference of the natural log of the spot rice prices to obtain the returns on spot rice, and we subtract the natural log of the spot rice prices from the natural log of the rice futures prices to calculate the rice futures premium.
\begin{center}
(Table \ref{fpe_table1} around here)
\end{center}

From Table \ref{fpe_table1}, we confirm that the longer the contract month, the more volatile the futures premium. Table \ref{fpe_table1} also shows the results of the unit root test with descriptive statistics for the data. For our estimations, all variables that appear in the moment conditions should be stationary. To confirm whether the variables satisfy the stationarity condition, we apply the ADF-GLS test of \citet{elliott1996eta}. We employ the modified Bayesian information criterion (MBIC) instead of the modified Akaike information criterion (MAIC) to select the optimal lag length. This is because, from the estimated coefficient of the detrended series, $\hat\psi$, we do not find the possibility of size-distortions (see \citet{elliott1996eta}; \citet{ng2001lls}).

%% file: fpe_empirical.tex
\section{Empirical Results}\label{fpe_sec5}
\subsection{Preliminary Results}
We assume a time-invariant regression model with constant parameters for our preliminary estimations. Table \ref{fpe_table2} summarizes our estimation results for a time-invariant regression model for the whole sample: all estimates for $\alpha$ in Equation (\ref{eqn:fpe eq}) are almost zero, and as the contract month of the rice futures is longer, the corresponding estimates of $\beta$ are larger. These results suggest that the longer the contract month, the more efficient the rice market in prewar Japan.\footnote{Note that our estimates of $\beta$ are slightly larger than \citetapos{taketoshi1999ear} estimates because our dataset is seasonally unadjusted.} \citet{taketoshi1999ear} obtains different estimates of $\beta$ for each subsample when the whole sample is split into four subsamples. However, the author does not examine whether the rice market in prewar Japan is efficient using conventional statistical inferences. Therefore, we use \citetapos{ito2016eme} non-Bayesian time-varying regression model to estimate $\beta$ because the rice market in prewar Japan may not always be efficient over time. However, we verify if the time-varying regression model is more appropriate than the time-invariant regression model before we adopt a new approach. Then, we apply \citetapos{hansen1992a} parameter constancy test to investigate whether the time-invariant model is a better fit for our data.
\begin{center}
(Table \ref{fpe_table2} around here)
\end{center}
Table \ref{fpe_table2} presents the results of our preliminary estimations: the estimates of time-invariant regression models ($\alpha$ and $\beta$) and their corresponding \citetapos{hansen1992a} joint parameter constancy test statistics ($L_C$). For our time-invariant regression model, the joint parameter constancy test rejects the null hypothesis of constancy at the 5\% level against the alternative hypothesis that the parameter variation follows a random walk process. These results suggest that the time-invariant regression model does not accommodate our data; rather, we should use the time-varying regression model for the prewar Japanese rice market data before World War I\hspace{-.1em}I.

\subsection{Time-varying Market Efficiency}
Figures \ref{fpe_fig1} and \ref{fpe_fig2} show that the farthest contract month transaction is more efficient than the nearest and the next-nearest months in both the Tokyo and Osaka rice futures markets.
\begin{center}
(Figures \ref{fpe_fig1} and \ref{fpe_fig2} around here)
\end{center}
The time-varying market efficiency measure of the farthest contract month shows a similar tendency as the next-nearest months. However, the longer the contract month (maturity) in futures, the more successfully traders hedged price risks in the rice market.
\begin{center}
(Figures \ref{fpe_fig3} and \ref{fpe_fig4} around here)
\end{center}

Figures \ref{fpe_fig3} and \ref{fpe_fig4} show that the farthest contract month transaction was larger than any other futures transaction in both Tokyo and Osaka. The farthest contract month transactions amounted to approximately 70\% of all futures in both rice markets. This suggests that the transaction successfully distributed required information to hedge price risks in the rice market in prewar Japan. Moreover, the time-varying market efficiency measure of the nearest contract month had a different nature than the next-nearest and farthest contract months in both Tokyo and Osaka (see Figures \ref{fpe_fig1} and \ref{fpe_fig2}). This interesting feature of the rice futures markets before World War I\hspace{-.1em}I in Japan reflects that the futures markets provided settlement on the balance and delivery of physical rice to clear the nearest month transactions.
\begin{center}
(Figure \ref{fpe_fig5} around here)
\end{center}

Figure \ref{fpe_fig5} shows the delivery volume for both Tokyo and Osaka. The figure indicates that the sellers constantly delivered physical rice to the buyers to clear their contracts in both cities. Taishichiro Tanaka, an administration officer of the Osaka Stock Exchange and a lecturer at Kobe College of Commerce, noted a characteristic of the nearest month transaction of rice in 1910. Tanaka stated, ``The nearest contract month transaction of rice has a property similar to the spot transaction.''\footnote{See \citet{tanaka1910sej} for details.} According to his assertion, the nearest contract month transaction had two aspects: a financial market for rice brokers to hedge price risks and a delivery space for physical rice. Consequently, the delivery space aspect reduced the market efficiency measure of the futures trading.

 Figures \ref{fpe_fig1} and \ref{fpe_fig2} show some periods when even the farthest contract month transaction in futures showed low market efficiency measure. Specifically, Figure \ref{fpe_fig1} exhibits three such periods for the Tokyo rice futures market: the late 1880s, the late 1890s, and from the mid-1900s to the 1910s; Figure \ref{fpe_fig2} exhibits three such periods for the Osaka rice futures market: the late 1880s, the late 1890s, and from the mid-1900s to the mid-1920s. The next section discusses why the farthest month transaction sometimes failed to hedge price risks in the rice market.

%% file: fpe_interpretation.tex
\section{Historical Interpretation}\label{fpe_sec6}
\subsection{Increasing Rice Exports from Japan in the Late 1880s}

In the late 1880s, the amount of rice exported increased and the Japanese government was directly involved in rice exporting. Specifically, the government bought the rice in the spot market, and sold it abroad as mentioned in Section \ref{fpe_sec1}. \citet[p.18]{omameuda1993fpm} argues that the growing rice exports from Japan in fact caused an increase in the spot rice price in the domestic market. In other words, government intervention in the rice spot market affected the price formation of rice. This situation reduced the efficiency measure. However, according to Figures \ref{fpe_fig1} and \ref{fpe_fig2}, the market efficiency measure in Tokyo was lower than in Osaka. This situation resulted from the difference in supply and demand between Eastern and Western Japan.

In the 1880s, the place of origin of rice circulated in Tokyo was different from in Osaka. In Tokyo and Osaka, the rice was supplied from Eastern and Western Japan, respectively. In addition, the rice produced in Chubu District located midway between Tokyo and Osaka was circulated in both cities (see \citet[pp.17--22]{mof1892psf}).\footnote{Chubu District includes Aichi, Shizuoka, Gifu, Nagano, Fukui, Ishikawa, Toyama, Niigata, and Yamanashi prefectures in the area surrounding Nagoya City.} That is, Tokyo and Osaka were the distribution and consumption centers for rice in Eastern and Western Japan, respectively. However, the rice supply in Eastern Japan was tighter than in Western Japan. Because rice is cultivated in tropical regions, it grows well in Western Japan where the climate is warmer than in Eastern Japan. For example, in 1888, the volumes of rice production per population in Eastern and Western Japan were 0.79 {\it koku} and 0.91 {\it koku}, respectively (see \citet[pp.45--105]{mac1890sya} and \citet[pp.2-12]{sb1909psp}). Under the rice supply situation, rice exports from Western Japan were brisk in the 1880s.

Rice wholesalers around the port of Kobe, the only trading port in Kinki District, were actively involved in rice exports in the 1880s. The wholesalers observed the change in domestic and export rice prices and chose either domestic retailers or foreign trade merchants as trading partners (see \citet[p.22]{omameuda1993fpm}). In contrast, rice wholesalers in Eastern Japan were not involved in rice exports. The Japanese government also did not buy the rice for export in Eastern Japan. The government bought the rice in Western Japan including Chubu District and exported it from ports in Western Japan such as the port of Kobe (see \citet[pp.162--163]{mof1919hrp}). In short, the government bought the rice in Western Japan and the Chubu District. Because the rice produced in Chubu District circulated in the tight rice market in Tokyo, the government intervention in the rice spot market in Chubu District seriously affected the price formation of rice in Tokyo. {\it Tokyo Asahi Shimbun} (Tokyo Asahi Newspaper) addressed this state of affairs in 1888 as follows.
\begin{quote}
``In recent days, the rice price in futures increased while the rice harvest is plentiful. The main reason for this situation is that the market participants expect rice exports from Japan to increase because of the poor wheat harvest in Europe.''\footnote{See \citet{asahi1888eir} for details.}
\end{quote}
In short, the government-led expansion of rice exports disrupted the desired relation between spot and futures prices. Accordingly, the government intervention reduced the market efficiency measure in Tokyo severely.

However, a population increase and high growth rates in the big cities such as Tokyo and Osaka resulted in greater rice consumption after Japan experienced a serious rice failure in 1890. Then, rice exports rapidly decreased. In the early 1890s, it was possible to hedge rice price risks in the Tokyo and Osaka rice futures markets.

\subsection{Increasing Rice Imports in the Late 1890s}

In contrast to the case of the late 1880s, rice imports grew rapidly in the late 1890s.
\begin{center}
(Figure \ref{fpe_fig6} around here)
\end{center}
Figure \ref{fpe_fig6} exhibits Japan’s rice consumption and imports from 1880 to 1932. The figure also shows that the ratio of imports to consumption began to increase from the 1890s. As mentioned in Section I, imported and domestic rice had differences in texture and taste. Accordingly, Japanese consumers considered imported rice an alternative grain. Consumers mixed domestic and imported rice in meal preparation (see \citet[pp.37--40]{omameuda1993fpm}). Both the Tokyo and Osaka rice futures markets listed only two standard brands of rice as domestic rice: Musashi (rice cropped in Saitama prefecture) and Settsu (rice cropped in Hyogo prefecture). \citet{ito2016meg} show that the most significant interventions responsible for the low market efficiency measures from the year 1890 were often government orders that the futures markets should use imported rice as an alternative to the listed domestic rice in futures transactions. That is, the commodity to be traded in the rice futures markets was domestic rice. When market participants perceived a significant difference for longer-term transactions, the futures market was irrelevant to the spot market.

However, the market efficiency measure estimated in this paper for the year 1890 is different from the estimation shown by \citet{ito2016meg}. The difference between the market efficiency measure estimated in this paper and \citet{ito2016meg} stem from the data used in the two papers. We include the futures market and the spot market while \citet{ito2016meg} use monthly rice futures data only to examine the efficiency measure of rice futures markets from 1881 to 1932. \citet{ito2016meg} focus on the futures market ignoring the increasing volume of actual imported rice in the spot market. According to Figure \ref{fpe_fig6}, the ratio of imported rice to rice consumption in 1890 was below 4\%. The ratio in 1890 was lower than the ratios in 1897 and 1898 when rice imports were higher again during the period after 1890. That is, in 1890, both the rice in the futures and spot markets was minimally different because imported rice was a minority portion of the physical rice distributed. Therefore, futures rice transactions in both Tokyo and Osaka were nearly efficient in 1890.

Figures \ref{fpe_fig1} and \ref{fpe_fig2} show that the efficiency measure of the Osaka rice futures market in 1890 was lower than that of the Tokyo rice futures market. This result was caused by critical confusion in the same year in the Osaka rice futures market. The Osaka rice futures market opposed the government's order in 1890 to amend the trading rules that allowed imported rice to be delivered as an alternative to domestic rice. The Ministry of Agriculture and Commerce revoked the permission for futures markets to trade in reaction to its disobedience. Consequently, futures trading in Osaka stopped temporarily, and this critical confusion reduced the market efficiency measure. The increasing volume of rice imports began to significantly reduce the efficiency measure of rice futures trading in the late 1890s. Figure \ref{fpe_fig6} shows that the ratio of import to consumption increased rapidly in the late 1890s because Japan was hit by a record poor harvest in 1897 (see \citet[p.184]{ota1938rpp}). In the Tokyo-Fukagawa rice spot market, the largest rice spot market in Japan, imported rice amounted to 43\% of all physical rice in 1898 (see \citet[pp.268--270]{sasaki1937htr}). In the same year, imported rice in year-end stock amounted to 54\% of all inventory in the Osaka rice spot market (see \citet[pp.100--102]{doujima1915hod}). These observations prove a significant difference in quality between rice traded in the futures market and the spot market. Imported rice was no longer the minority portion of rice distribution in the late 1890s. The rice market efficiency measure reduced in this situation. At that time, the rice market in Japan traded in different grain because only domestic rice was listed in the futures market, and both imported and domestic rice were traded in the spot market. Although the government forced the futures markets to accept the imported rice as an alternative to the listed deliverable domestic rice temporarily from January to October in 1898, the futures price failed to be an accurate index of the expected price of rice in the late 1890s including 1898. In August 1898, {\it Yomiuri Shimbun} (Yomiuri Newspaper) addressed this situation as follows.
\begin{quote}
``Since the exchanges have agreed to deliver imported rice, the rice futures market depends on price fluctuations for imported rice. As a result, the trend in rice futures price was different from that of the spot price.''\footnote{See \citet{yomiuri1898cad} for detail.}
\end{quote}

This situation reflected that the grading system for delivered imported rice was different from the domestic rice system in the futures markets. The futures markets set up the substandard imported rice and changed the price difference between the grading of the substandard rice and imported rice. On the other hand, the price difference between standard domestic rice and substandard imported rice was basically fixed while the price difference between domestic and imported rice in the spot market was flexible. Yoshinari Kawai, a bureaucrat of the Ministry of Agriculture and Commerce from 1911 to 1919, stated the reason why the futures markets set up the above-mentioned grading system (see \citet[p.81]{hata1981scp}).
\begin{quote}
``The quality of Korean and Taiwanese rice is lower than that of domestic rice. In the Tokyo futures market, the price gap between Korean rice and standard domestic rice is from three yen to four yen. The price gap between Taiwanese rice and standard domestic one is from five yen to six yen. However, the lowest quality of deliverable domestic rice is priced about two yen less than the standard rice. This price gap is smaller than the gap in price between imported rice and standard domestic rice. Since the quality difference between imported rice and standard domestic rice is too large, nobody can value the gap between imported rice and domestic rice. This situation obliges the futures market to set up the substandard imported rice.''\footnote{See \citet[pp.297--298]{kawai1921loe} for detail.}
\end{quote}

In summary, because the quality of imported rice was not the same as domestic rice, the futures market could not value the gap in price between the two. As a result, this price gap in the futures market could not move with the gap in the spot market, and the market structure of rice futures differed from the structure of the rice spot market. This dislocation in the rice futures markets was settled at the end of the 1890s. In November 1898, the government repealed the order for the futures markets to accept the imported rice as an alternative to the listed domestic rice. In addition, the ratio of rice imports to rice consumption decreased rapidly in 1899. However, this ratio began to skyrocket in 1902 and remained above approximately 5\%. Japan had been a continuous importer of rice since the 1890s.

\subsection{Difference in Structure Between the Spot and Futures Markets from the Mid-1900s to the Mid-1920s.}

We discuss the decrease in the efficiency measure from the mid-1900s to the mid-1920s. The market efficiency measure in both Tokyo and Osaka began to decrease in the mid-1900s. Tokyo Asahi Shimbun reported on the rice market in the late 1900s as follows.

\begin{quote}
 ``The price of imported rice was higher than coarse cereals, the price of domestic rice in the spot market was higher than imported rice, and the price of domestic rice in the futures market was higher than domestic rice in the spot market. The futures price was highest in the rice and grains market because most of the participants in the exchanges were speculators.''\footnote{See \citet{asahi1907rpr} for details.}
\end{quote}

That is, the price difference between the spot and the futures markets was caused by speculative rice trader dealings in the futures markets. However, there are two discrepancies in the low efficiency measures between Tokyo and Osaka from the mid-1900s to the mid-1920s. First, the Osaka rice futures market failed to provide an accurate index of the expected price of spot rice that was superior to Tokyo. Second, the period of market inefficiency in the Osaka rice futures market was longer than that of Tokyo. The futures in the Tokyo rice futures market failed to provide a fine index of the expected price of spot rice from the mid-1900s to the 1910s. The efficiency measure of the Osaka rice futures market reduced until the mid-1920s. The difference in the market efficiency measures between Tokyo and Osaka was a consequence of the diversity in market structure characterized by the distribution of imported rice.

According to Figure \ref{fpe_fig6}, the volume of Japanese rice imports remained at a low level from 1899 to 1901, although that volume began to increase in 1902. Specifically, the ratio of imported rice to all physical rice arrived from 1901 to 1905 was 27\% in the Tokyo-Fukagawa rice spot market and 23\% in the Osaka rice spot market (see \citet[pp.72--73]{osaka1903ocs}; \citet[pp.79--80]{osaka1906ocs}; \citet[pp.87--88]{osaka1907ocs}; \citet[pp.268--270]{sasaki1937htr}). Consequently, the market efficiency measure in both Tokyo and Osaka decreased in the early 1900s because the market structure of rice futures was different from the structure of the rice spot market (see Figures \ref{fpe_fig1} and \ref{fpe_fig2}). This situation changed in the late 1900s.

The volume of Japanese rice imports in Japan began to decrease beginning in 1906 (see Figure \ref{fpe_fig6}). From 1905 to 1909, the ratio of imported rice to all physical rice arrived in Tokyo decreased to 22\% while this ratio in Osaka increased to 27\% (see \citet[pp.91--92]{osaka1908ocs}; \citet[pp.163--164]{osaka1909ocs}; \citet[pp.163--164]{osaka1910ocs}; \citet[pp.253--254]{osaka1911ocs}; \citet[pp.268--270]{sasaki1937htr}).\footnote{The ratio of imported rice to all physical rice arrived in the Osaka rice spot market does not include the data for 1902, 1903, and 1908 because there are no statistics.} This asymmetric situation between Tokyo and Osaka was caused by an increase in imported rice from Taiwan and Korea. From 1905 to 1909, the import volume of Taiwanese rice increased from 0.7 million {\it koku} to 1.2 million {\it koku}, and the import volume of Korean rice increased from 0.1 million {\it koku} to 0.5 million {\it koku} while the volume of foreign rice imports decreased from 2.6 million {\it koku} to 0.9 million {\it koku} (see \citet[pp.3--4]{maf1933dor}). To sum up rice imports in the late 1900s, the volume of all rice imports decreased along with the import volume of foreign rice, whereas Taiwanese and Korean rice imports increased.

The major destination port for Taiwanese and Korean rice in Japan was the port of Kobe. When the rice from Taiwan and Korea arrived at Kobe, barge clusters transported Taiwanese and Korean rice to the nearby port of Osaka, which was almost nine miles (15 kilometers) away (see \citet[p.561]{ichinoue1920ssr}). Taiwanese and Korean rice differed in quality from domestic rice. Taiwanese and Korean rice was categorized as the indica variety as was the imported rice from Indochina. Taiwanese rice varied considerably in quality because the rice farming method in Taiwan was double cropping (see \citet[p.27]{maf1938rtf}). Korean rice was mixed with sand, stone, and other impurities (see \citet[p.598]{hishimoto1938skr}). Korean rice was concentrated in Osaka because rice traders who could decontaminate the impurities from Korean rice clustered in Osaka (see \citet[p.2]{maf1938rtf}). Osaka Asahi Shimbun reported, ``In Osaka, rice traders have excellent sales for Taiwanese rice because its demand has increased.''\footnote{See \citet{asahi1909rft} for details.}

However, rice futures markets could not trade the imported rice. In the late 1900s, both the standard and deliverable commodities in the futures markets were domestic rice only. Therefore, the structure of the futures market differed from the spot market. In the late 1900s, the market efficiency measure reduced only in Osaka because the ratio of imported rice to all physical rice arrived in Osaka increased. As a practical measure, Tokyo Asahi Shimbun urged the government to shrink the market structure differences between the spot and the futures.\footnote{See \citet{asahi1911tst} for details.} Although the government forced the futures markets to accept imported rice from Taiwan and Korea as an alternative to deliverable domestic rice in June 1912, the futures price in both Tokyo and Osaka failed to be an accurate index of the expected price in the rice spot market in the 1910s.

The inefficiency resulted from the persistent differences in structure between the spot and the futures markets. A greater amount of imported rice had been traded in the spot market since the 1910s. From 1903 to 1918, the volume of domestic rice production increased to 17.7\%. In the same period, the total population and the urban population with over 100,000 inhabitants increased to 20.2\% and 63.1\%, respectively (see \citet[pp.12, 14, 108]{boj1966hys}). Japan faced a rice shortage because its population growth was greater than its increase in rice production. Japan was forced to compensate for the rice shortage with rice imports. In the 1910s, the ratio of import volume to consumption fluctuated dramatically. The ratio increased sharply from 1912 to 1913 and from 1917 to 1919 (see Figure \ref{fpe_fig6}).
\begin{center}
(Figures \ref{fpe_fig7} and \ref{fpe_fig8} around here)
\end{center}

The ratio of imported rice to all physical rice arrived in both Tokyo and Osaka had a similar tendency as the ratio of import volume to consumption. However, the ratio of imported rice to all physical rice arrived in Osaka showed a different propensity to that of Tokyo in the early 1910s. The ratio of imported rice to all physical rice arrived in Osaka continued to increase to approximately 50\% until 1915 while the rate in Tokyo reached a peak in 1913 and remained approximately 40\% until 1915. This asymmetric trend between Tokyo and Osaka was caused by an increase in the volume of Korean rice imports under the same conditions in the late 1900s. From 1912 to 1915, the volume of Korean rice imports increased from 0.2 million {\it koku} to 19 million {\it koku}, and the volume of foreign rice imports decreased from 20 million {\it koku} to 0.5 million {\it koku} (see \citet[pp.3--4]{maf1933dor}). The increase in the volume of Korean rice imports led to a swelling in the volume of rice arrived from overseas in Osaka. In Osaka prefecture, the volume of imported rice arrived increased from 0.2 million {\it koku} in 1912 to 0.9 million {\it koku} in 1915 with an increase in the volume of Korean rice arrived from 0.2 million {\it koku} to 0.7 million {\it koku} for the same period (see \citet[p.420]{ichinoue1920ssr}).

In the late 1910s, the market efficiency measure temporarily improved from 1916 to 1917. This was because the difference in market structure between the spot and the futures had shrunk along with the ratio of imported rice to all physical rice arrived in both Tokyo and Osaka (see Figures \ref{fpe_fig1}, \ref{fpe_fig2}, \ref{fpe_fig7}, and \ref{fpe_fig8}). Subsequently, Japan faced a serious rice shortage from 1918 to 1919. The nationwide riots, {\it kome-soudo} (the rice riots) in 1918, were a response to the shortage. At the same time, the ratio of imported rice to all physical rice arrived in both Tokyo and Osaka surged. From 1912, the government forced the utures markets to accept imported rice from Taiwan and Korea as an alternative to listed domestic rice. However, the differences between the spot and futures markets remained.

Yoshinari Kawai, Director of the Division of Foreign Rice Management at the Ministry of Agriculture and Commerce from 1918 to 1920 and responsible for controlling the rice market to stabilize prices and transactions, stated that the fixed price difference in the futures markets between the standard domestic rice and substandard imported rice was smaller than the flexible price difference in the spot market between domestic and imported rice (see \citet[p.201]{maf1981hyh}). Therefore, rice traders considered Korean and Taiwanese rice more desirable for delivery in the rice futures markets than sale in the spot markets. In 1921, Kawai reviewed futures trading in the 1910s as follows:

\begin{quote}
 ``At the maturity date in the exchanges, many sellers tended to use Taiwanese rice as the rice for delivery. As a result, the futures price fell. However, the price difference between the spot and the futures did not shrink at the maturity date. The futures market showed hardly any relationship with the spot market because the rice futures market strongly depended on the price of rice from Taiwan and Korea.''\footnote{See \citet[pp.299--300]{kawai1921loe} for details.}
\end{quote}
In the 1910s, the differences in structure between the spot and the futures markets remained because of the failure of the imported rice grading system in the futures markets. Consequently, the efficiency measure decreased in the same period. This situation suggested that rice traders could not hedge price volatility risk in the domestic rice market.　

In the early 1920s, the turmoil in the rice markets settled. In the same period, the volume of imported rice arrived in Tokyo decreased and the ratio of imported rice to all physical rice arrived remained below 30\% (see Figure \ref{fpe_fig7}). For Osaka, the major destination of Taiwanese and Korean rice, the volume of imported rice arrived increased, and the ratio of imported rice to all physical rice arrived remained over 50\% (see Figure \ref{fpe_fig8}). Asahi Shimbun reported that the Osaka rice futures market failed to provide an accurate index of the expected price in the spot market in the early 1920s. The newspaper stated, ``The futures market in Osaka was disrupted because 440,000 {\it koku} of Korean rice was stored in the city of Osaka. Additionally, the large amount of Korean rice was stored around Osaka.''\footnote{See \citet{asahi1924ikr} for details.}

Therefore, while the market efficiency measure improved in Tokyo, it decreased in Osaka, reflecting the asymmetric situation between Tokyo and Osaka.

\subsection{Improving Market Efficiency in the Late 1920s}

In the late 1920s, Tokyo and Osaka futures provided an accurate index of the expected price of spot rice. The reasons for the improvement in the efficiency measure were two-fold. First, the rice cultivated in Taiwan and Korea changed in quality. Second, government-owned physical rice increased after the late 1920s. With respect to the changes in Taiwanese and Korean rice, a newspaper published in June 1924 reported that the quality differences between domestic and Korean rice were less pronounced.

\begin{quote}
``On June 18, the Society of Housewives held a rice tasting event in Tokyo. About 70 people from the Tokyo Metropolitan Government, the Ministry of Agriculture and Commerce, the Bank of Korea, the Oriental Development Company (Toyo Takushoku Company), and so on were invited to the event. The participants tasted the domestic rice and the Korean rice. They tried to guess which Korean rice was in all the samples. However, only five of them, who were the experts on rice trading, guessed Korean rice exactly.''\footnote{See \citet{asahi1924brd} for details.}
\end{quote}

After the 1910s, the Governor-General of Korea promoted rice cropping of the japonica rice variety. The ratio of domestic rice to total rice production in Korea was only 5\% in 1912, 69\% in 1921, and 79\% in 1932 (see \citet[pp.438--439]{tobata1939rek}). Additionally, following Korea, the Governor-General of Taiwan introduced japonica rice cropping in 1924 (see \citet[p.244]{taiwan1945art}). Since the introduction of japonica rice cropping in Korea and Taiwan, rice cropped in the Japanese colonies had represented most imported rice in Japan. From 1920 to 1930, the changes in Korean, Taiwanese, and foreign rice import volumes and their ratios to total rice imports were as follows: the import volume of rice from Korea increased from 1.7 million {\it koku} to 5.2 million {\it koku}, Taiwanese rice import volume increased from 0.7 million {\it koku} to 2.2 million {\it koku}, and foreign rice import volume increased from 0.8 million {\it koku} to 1.2 million {\it koku}. Each change in ratio was from 54.4\% to 60.1\%, from 21.8\% to 25.4\%, and from 24.7\% to 14.5\% in the same order (see \citet[pp.4--5]{maf1933dor}). This implies that the proportion of imported rice categorized as the indica variety arrived in the rice spot market rapidly decreased. Thus, imported rice had a minimal effect on rice pricing in the rice futures market and the spot market even under conditions of a large amount of imported rice. The futures markets shrank the price difference in the grading system between substandard imported rice and standard domestic rice delivery after 1925.\footnote{See \citet{asahi1924mag} for details.}

In summary, the changes in rice variety cropped in Korea and Taiwan allowed inland populations to consume more imported rice; the changes also led to better hedging of price risk in the rice futures markets.

We discuss the second factor in the efficiency measure improvements in the late 1920s. With respect to the establishment of the Rice Law in 1921 and its amendment in 1925 by the government, the amended law allowed the government to buy and sell physical rice directly to adjust the circulation volume and the rice price. Specifically, the Ministry of Agriculture and Forestry had owned the physical rice in stock since 1925 to adjust the volume of circulation in the rice market. The stock rice owned by the government amounted to one million {\it koku} in 1925 and reached 2.3 million {\it koku} in 1931 (see \citet[pp.4--5]{maf1935tvp3}). This volume in 1931 accounted for 3.5\% of total rice demand in Japan and equaled nearly 44\% of the annual volume of rice arrived in Tokyo (see Figures \ref{fpe_fig6} and \ref{fpe_fig7}). From the mid-1920s, government intervention based on the Rice Law and the change in the variety of imported rice from Korea and Taiwan stabilized the supply and demand of rice in the main islands of Japan. This implies that the price risk in the rice market decreased. However, the seasonality price risk remained. According to Figures \ref{fpe_fig3} and \ref{fpe_fig4}, trading volume in both the Tokyo and Osaka rice futures markets experienced seasonal fluctuations from the mid-1920s. The trading volume increased in harmony with the expansion of rice distribution in the fall harvest seasons from the mid-1920s. Therefore, market efficiency measures in both Tokyo and Osaka improved after the mid-1920s. However, the market efficiency measure was relatively lower in Osaka than in Tokyo. The difference in market efficiency measure between Tokyo and Osaka reflected the difference in their market structures.

After the late 1920s, the ratio of imported rice to all physical rice arrived in Osaka was approximately 80\% while the corresponding ratio to Tokyo was approximately 30\% (see Figures \ref{fpe_fig7} and \ref{fpe_fig8}). In short, imported rice had occupied the central position in the rice spot market in Osaka in contrast to Tokyo. Korean rice, which accounted for a large part of imported rice in Osaka, was of the same variety as domestic rice, but the price of rice cropped in Korea was lower than domestic rice. For example, the prices per {\it koku} of domestic and Korean rice in Osaka in 1931 were 18.15 yen and 17.29 yen, respectively. That is, consumers did not consider Korean rice to be the equivalent of domestic rice. This suggests that the differences in market structure between the spot and the futures markets reduced the market efficiency measure.

%% file: fpe_conclusion.tex
\section{Conclusion}\label{fpe_sec7}

This paper argues that a difference in market structure between spot and futures caused the functional decline of rice futures markets that were supposed to provide efficient forecasts of the rice spot price in prewar Japan. This paper also argues that government intervention in the markets failed to offset the decline. In contrast, the Japanese colonial government changed in the variety of colonial rice, and the function of rice futures was improved. Particularly, the difference was caused by rice futures markets' exclusive activity in domestic rice, whereas imported rice of a different variety represented a greater proportion of rice distribution in Japan.

The government intervened in rice futures exchanges to reduce the difference. The government forced the major exchanges in Tokyo and Osaka to allow the imported rice to be delivered, and the intervention practically fixed the differences between the prices of domestic and imported rice in the futures markets. The corresponding difference in the spot rice markets could not be adjusted and failed to improve the functioning of the rice futures markets.

However, when the Japanese colonial government promoted japonica rice cropping in the 1910s, market improvements occurred. The promotion reduced the differences in quality between domestic and imported (Korean and Taiwanese) rice that was distributed in Japan and the differences in structure between spot and futures markets. Finally, the rice futures markets regained their functioning.

In summary, the increased volume of imported rice of a variety different from domestic rice first disrupted the rice futures in prewar Japan. Then, government intervention in the rice futures markets failed to improve the disruption. Changes in colonial rice cropping successfully improved the disruption, and colonial rice was promoted to unify the different varieties of inland and colonial rice. The function of rice futures markets crucially depended on the differences in rice spot markets' structure.

%% file: fpe_ack.tex
\section*{Acknowledgments}

We would like to thank the editor, Jaime Reis, three anonymous referees, Shigehiko Ioku, Makoto Kasuya, Munechika Katayama, Yo Kikkawa, Kozo Kiyota, Junsoo Lee, Kris Mitchener, Tamaki Miyauchi, Chiaki Moriguchi, Toshihiro Nagahiro, Tetsuji Okazaki, Minoru Omameuda, Mototsugu Shintani, Masato Shizume, Yasuo Takatsuki, Masahiro Uemura, Tatsuma Wada, Asobu Yanagisawa, seminar participants at Doshisha University, Keio University, and Wakayama University, and conference participants at the Japanese Economics Association 2014 Autumn Meeting, and the 90th Annual Conference of the Western Economic Association International for their helpful comments and suggestions. We also acknowledge the financial assistance provided by the Japan Society for the Promotion of Science Grant-inAid for Scientific Research Nos. 26380397 (Mikio Ito), 26780199 (Kiyotaka Maeda), and 15K03542 (Akihiko Noda). All data and programs used for this paper are available on request.

%% file: fpe_table.tex
\setcounter{table}{0}
\renewcommand{\thetable}{\arabic{table}}

\clearpage

\begin{landscape}
\begin{table}[tbp]
\caption{Descriptive Statistics and Unit Root Tests}\label{fpe_table1}
\begin{center}
\scriptsize
\begin{tabular}{lllrrrrrrrrrrrrrr}\hline\hline
 &  &  & \multicolumn{6}{c}{Tokyo} &  & \multicolumn{6}{c}{Osaka} & \\\cline{4-9}\cline{11-16}
 &  &  & \multicolumn{2}{c}{One Month} & \multicolumn{2}{c}{Two Month} & \multicolumn{2}{c}{Three Month} &  & \multicolumn{2}{c}{One Month} & \multicolumn{2}{c}{Two Month} & \multicolumn{2}{c}{Three Month} & \\\cline{4-16}
 &  &  & \multicolumn{1}{c}{SR} & \multicolumn{1}{c}{FP} & \multicolumn{1}{c}{SR} & \multicolumn{1}{c}{FP} & \multicolumn{1}{c}{SR} & \multicolumn{1}{c}{FP} &  & \multicolumn{1}{c}{SR} & \multicolumn{1}{c}{FP} & \multicolumn{1}{c}{SR} & \multicolumn{1}{c}{FP} & \multicolumn{1}{c}{SR} & \multicolumn{1}{c}{FP} & \\\hline
 & Mean &  & 0.0025  & $-0.0172$  & 0.0015  & $-0.0172$  & 0.0024  & $-0.0176$  &  & 0.0013  & $-0.0227$  & 0.0027 & $-0.0212$ & 0.0042 & $-0.0201$ & \\
 & SD &  & 0.0536  & 0.0688  & 0.0908  & 0.0822  & 0.1145  & 0.0902  &  & 0.0619  & 0.0789  & 0.0923  & 0.0920  & 0.1171  & 0.1045 & \\
 & Min &  & $-0.3670$  & $-0.3623$ & $-0.4601$ & $-0.3600$ & $-0.5360$ & $-0.3549$ &  & $-0.2939$ & $-0.3780$ & $-0.3821$ & $-0.5279$ & $-0.4582$ & $-0.5991$ & \\
 & Max &  & 0.2791 & 0.1769 & 0.3116 & 0.2037 & 0.4752 & 0.1618 &  & 0.6569 & 0.2190 & 0.7260 & 0.3132 & 0.8131 & 0.2494 & \\\hline
 & ADF-GLS &  & $-16.1633$ & $-6.9630$ & $-4.2744$ & $-5.8867$ & $-6.6164$ & $-7.9606$ &  & $-13.8955$ & $-8.8637$ & $-5.7255$ & $-7.9995$ & $-4.8899$ & $-7.5746$ & \\
 & Lags &  & 0 & 0 & 8 & 0 & 10 & 1 &  & 1 & 0 & 8 & 1 & 2 & 18 & \\
 & $\hat\phi$ &  & 0.3442 & 0.8317 & 0.7294 & 0.8444 & 0.7843 & 0.8629 &  & 0.1968 & 0.7739 & 0.6053 & 0.7220 & 0.7153 & 0.7657 & \\\hline
 & $\mathcal{N}$ &  & \multicolumn{2}{c}{538} & \multicolumn{2}{c}{417} & \multicolumn{2}{c}{624} &  & \multicolumn{2}{c}{619} & \multicolumn{2}{c}{618} & \multicolumn{2}{c}{617} & \\\hline\hline
\end{tabular}
\vspace*{5pt}
{
\begin{minipage}{600pt}
\scriptsize
{\underline{Notes:}}
\begin{itemize}
\item[(1)] ``SR'' denotes the spot returns ($\log{S_{t+k}}-\log{S_t}$),
	    and ``FP'' denotes the futures premiums ($\log{F_{t+k|t}}-\log{S_t}$).
\item[(2)] ``ADF-GLS'' denotes the ADF-GLS test statistics, ``Lags'' 
           denotes the lag order selected by the MBIC, and ``$\hat\phi$'' 
           denotes the coefficients vector in the GLS detrended series 
           (see Equation (6) in \citet{ng2001lls}).
\item[(3)] In computing the ADF-GLS test, a model with a time trend 
           and a constant is assumed. The critical value at the 1\% 
           significance level for the ADF-GLS test is ``$-3.42$''.
\item[(4)] ``$\mathcal{N}$'' denotes the number of observations.
\item[(5)] R version 3.3.2 was used to compute the statistics.
 \end{itemize}
\end{minipage}}%
\end{center}
\end{table}
\end{landscape}

\clearpage

\begin{landscape}
\begin{table}[tbp]
\caption{Time-invariant Estimations}\label{fpe_table2}
\begin{center}
\begin{tabular}{clccccccccc}\hline\hline
 &  &  & \multicolumn{3}{c}{Tokyo} & & \multicolumn{3}{c}{Osaka} & \\\cline{4-6}\cline{8-10}
 &  &  & One Month & Two Month & Three Month & & One Month & Two Month & Three Month & \\\hline
 & \multirow{2}*{$\alpha$} & & 0.0056 & 0.0066 & 0.0118 & & 0.0076 & 0.0100 & 0.0119 & \\
 &  &  & [0.0028] & [0.0096] & [0.0085] & & [0.0030] & [0.0058] & [0.0106] & \\
 & \multirow{2}*{$\beta$} & & 0.1771 & 0.2927 & 0.5298 &  & 0.2771 & 0.3424 & 0.3823 & \\
 &  &  & [0.0743] & [0.1682] & [0.1572] & & [0.0435] & [0.0693] & [0.1023] & \\\hline
 & $\bar{R}^2$ &  & 0.0500 & 0.0679 & 0.1728 &  & 0.1232 & 0.1150 & 0.1149 & \\
 & $L_C$ &  & 1.0740  & 1.0510 & 1.6171 & & 1.5787 & 1.8331 & 2.4301 & \\\hline\hline
\end{tabular}
\vspace*{5pt}
{
\begin{minipage}{550pt}
\footnotesize
{\underline{Notes:}}
\begin{itemize}
\item[(1)] ``${\bar{R}}^2$'' denotes the adjusted $R^2$, and ``$L_C$'' denotes \citetapos{hansen1992a} joint $L$ statistic with variance.
\item[(2)] \citetapos{newey1987sps} robust standard errors are in brackets.
\item[(3)] R version 3.3.2 was used to compute the estimates and the statistics.
\end{itemize}
\end{minipage}}%
\end{center}
\end{table}
\end{landscape}

\clearpage

\begin{figure}[bp]
 \caption{Time-varying Estimates of $\beta$: The Case of the Tokyo Rice Market}\label{fpe_fig1}
 \begin{center}
 \includegraphics[clip,height=6.5cm,width=9.5cm]{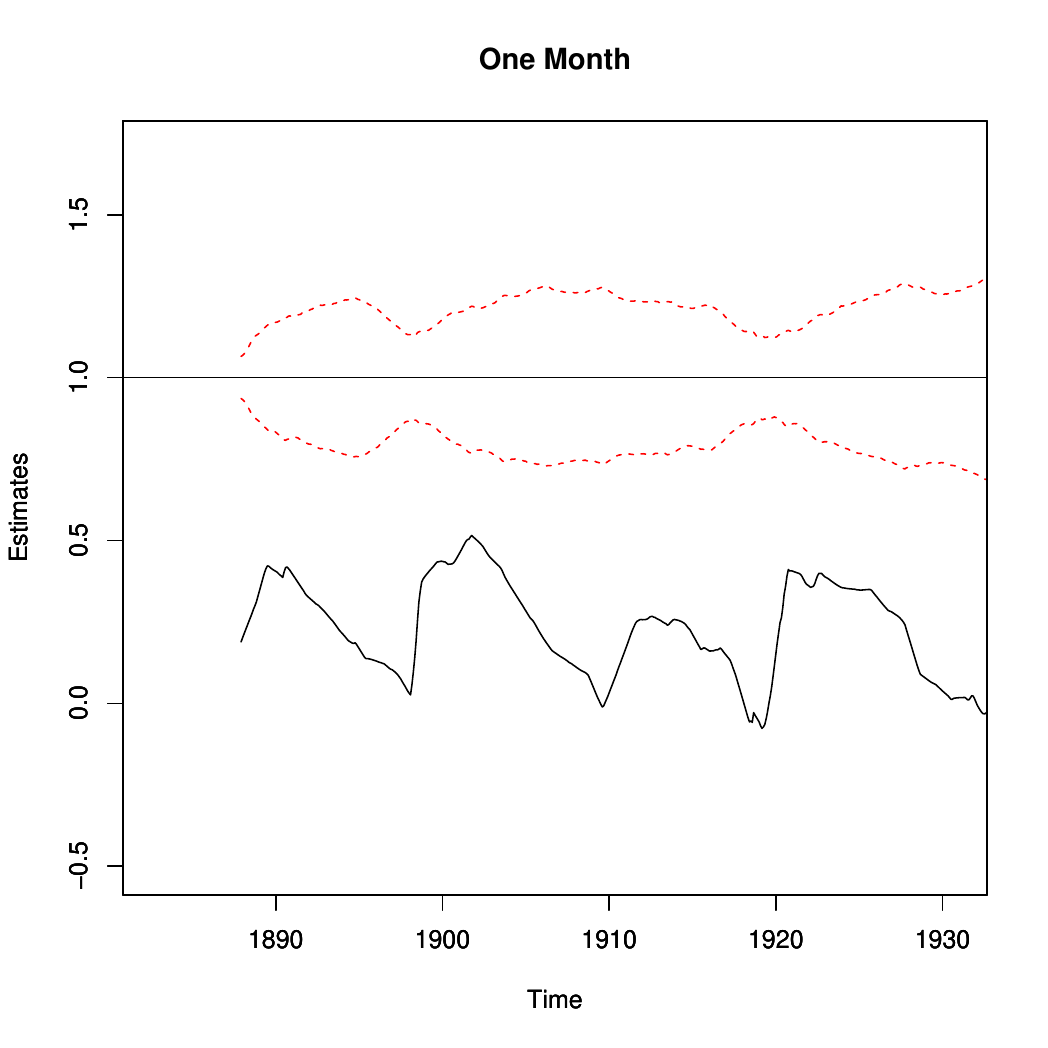}\\
 \includegraphics[clip,height=6.5cm,width=9.5cm]{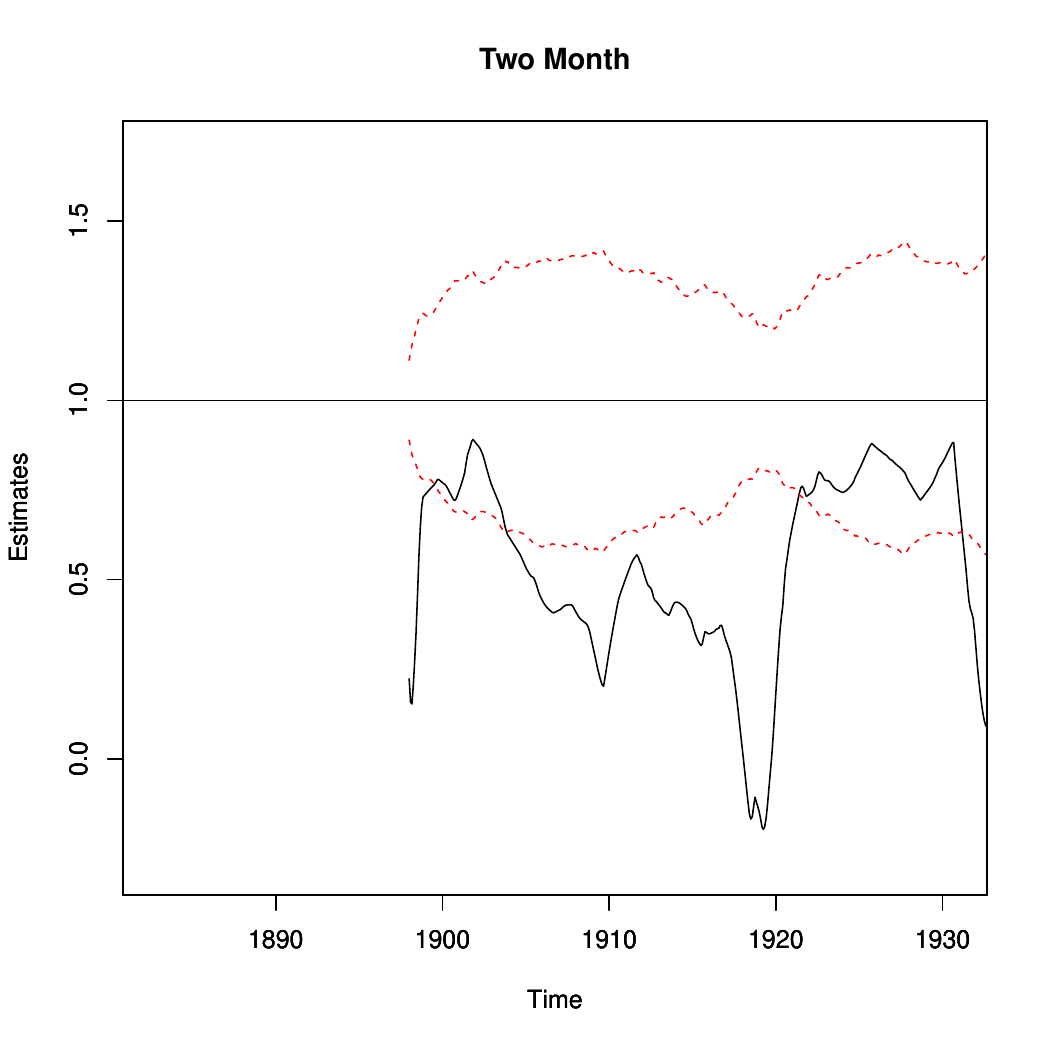}\\
 \includegraphics[clip,height=6.5cm,width=9.5cm]{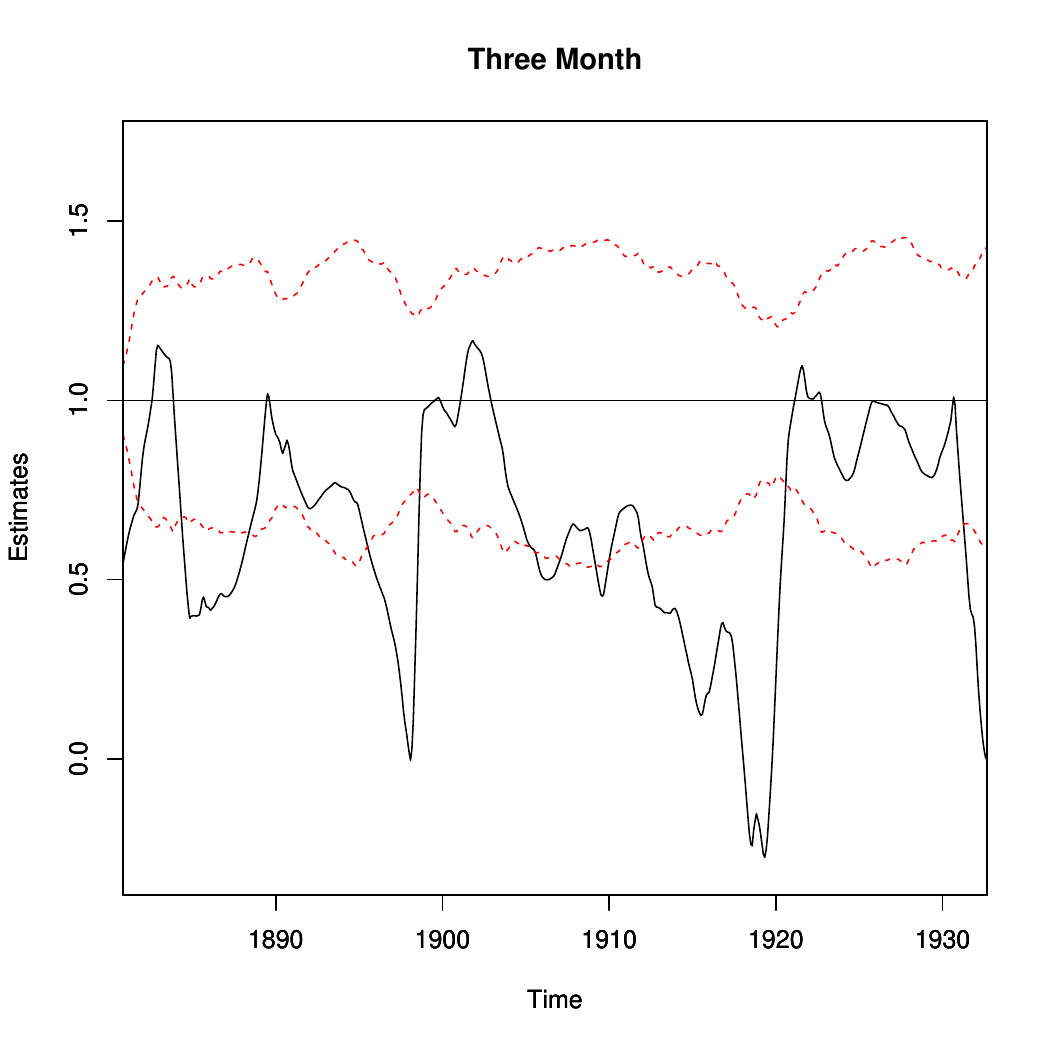}
\vspace*{3pt}
{
\begin{minipage}{650pt}
\scriptsize
\underline{Notes}:
\begin{itemize}
\item[(1)] The dashed red lines represent the 95 percent confidence bands of the estimates in the case of an efficient market. 
\item[(2)] We run 5,000 times bootstrap sampling to calculate the confidence bands.
\item[(3)] R version 3.3.2 was used to compute the estimates.
\end{itemize}
\end{minipage}}%
\end{center}
\end{figure}

\clearpage

\begin{figure}[bp]
 \caption{Time-varying Estimates of $\beta$: The Case of the Osaka Rice Market}\label{fpe_fig2}
 \begin{center}
 \includegraphics[clip,height=6.5cm,width=9.5cm]{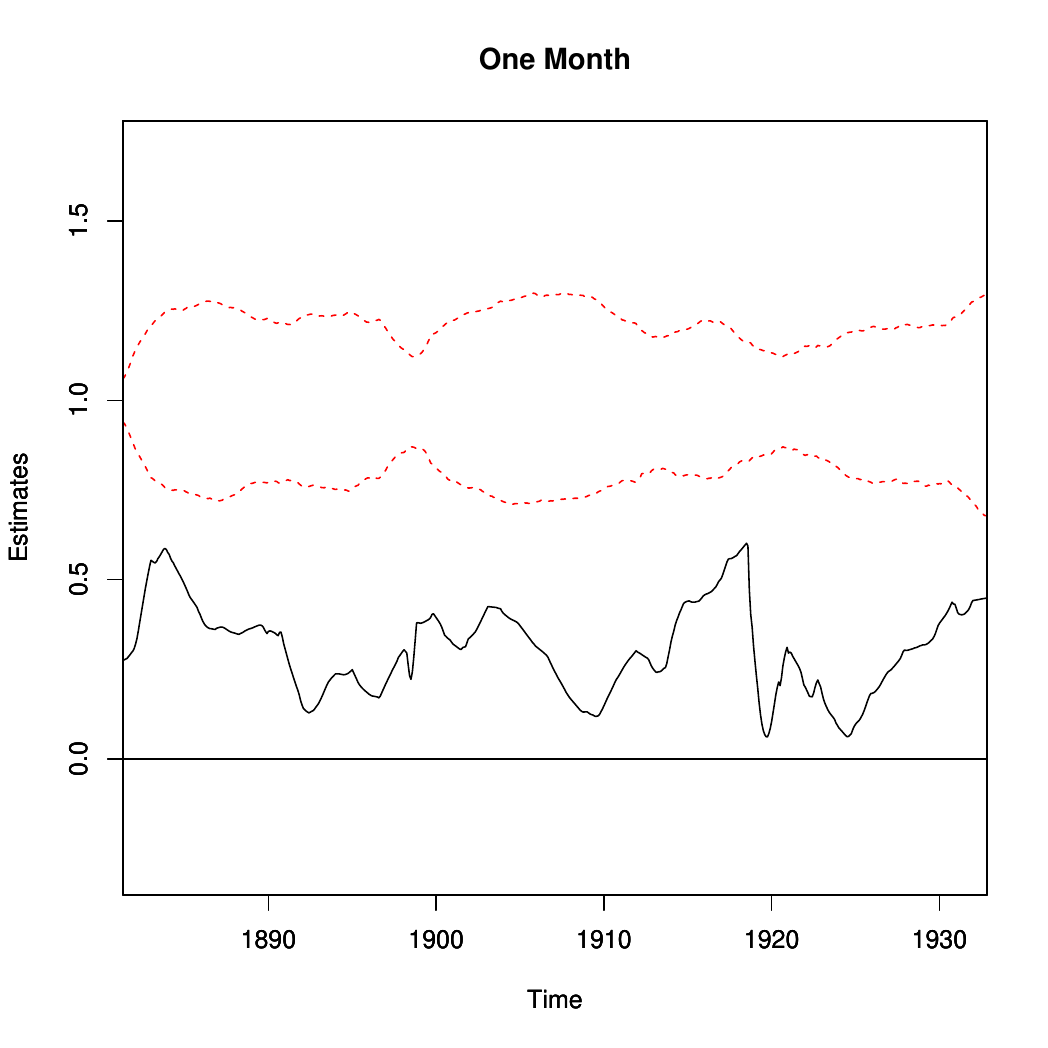}\\
 \includegraphics[clip,height=6.5cm,width=9.5cm]{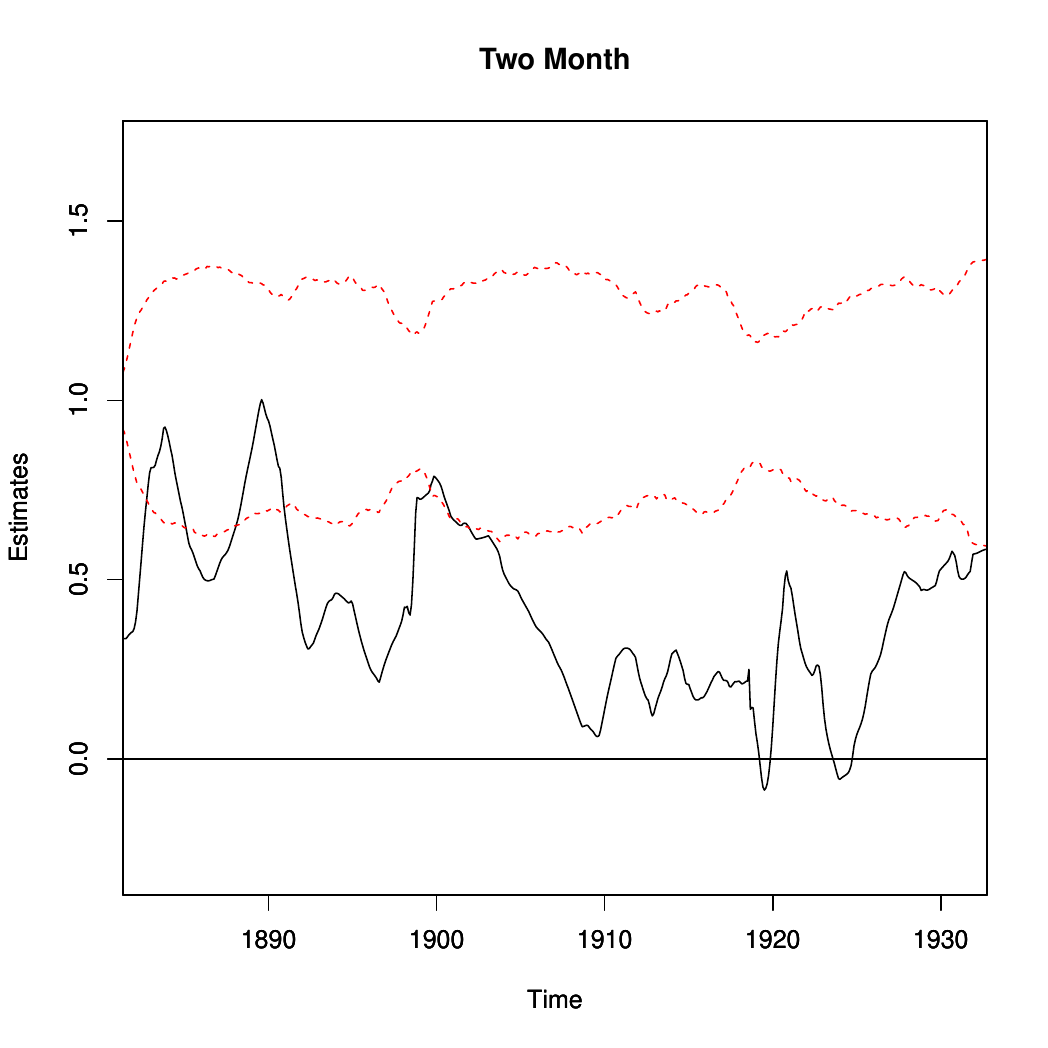}\\
 \includegraphics[clip,height=6.5cm,width=9.5cm]{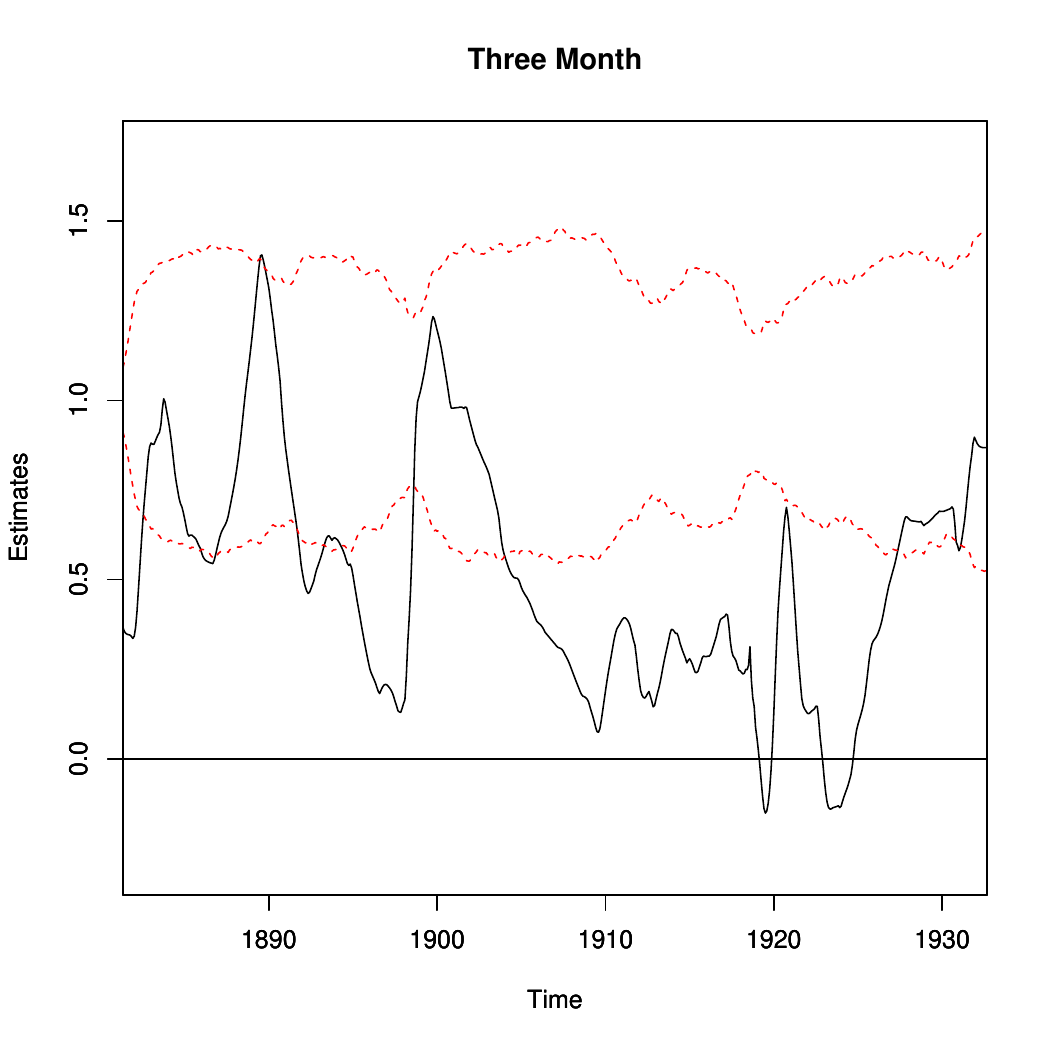}
\vspace*{3pt}
{
\begin{minipage}{650pt}
\footnotesize
\underline{Note}: As for Figure \ref{fpe_fig1}.
\end{minipage}}%
  \end{center}
\end{figure}

\clearpage

\begin{figure}[bp]
 \caption{Trading Volume of Rice in the Tokyo Rice Exchange}\label{fpe_fig3}
 \begin{center}
 \includegraphics[scale=0.9]{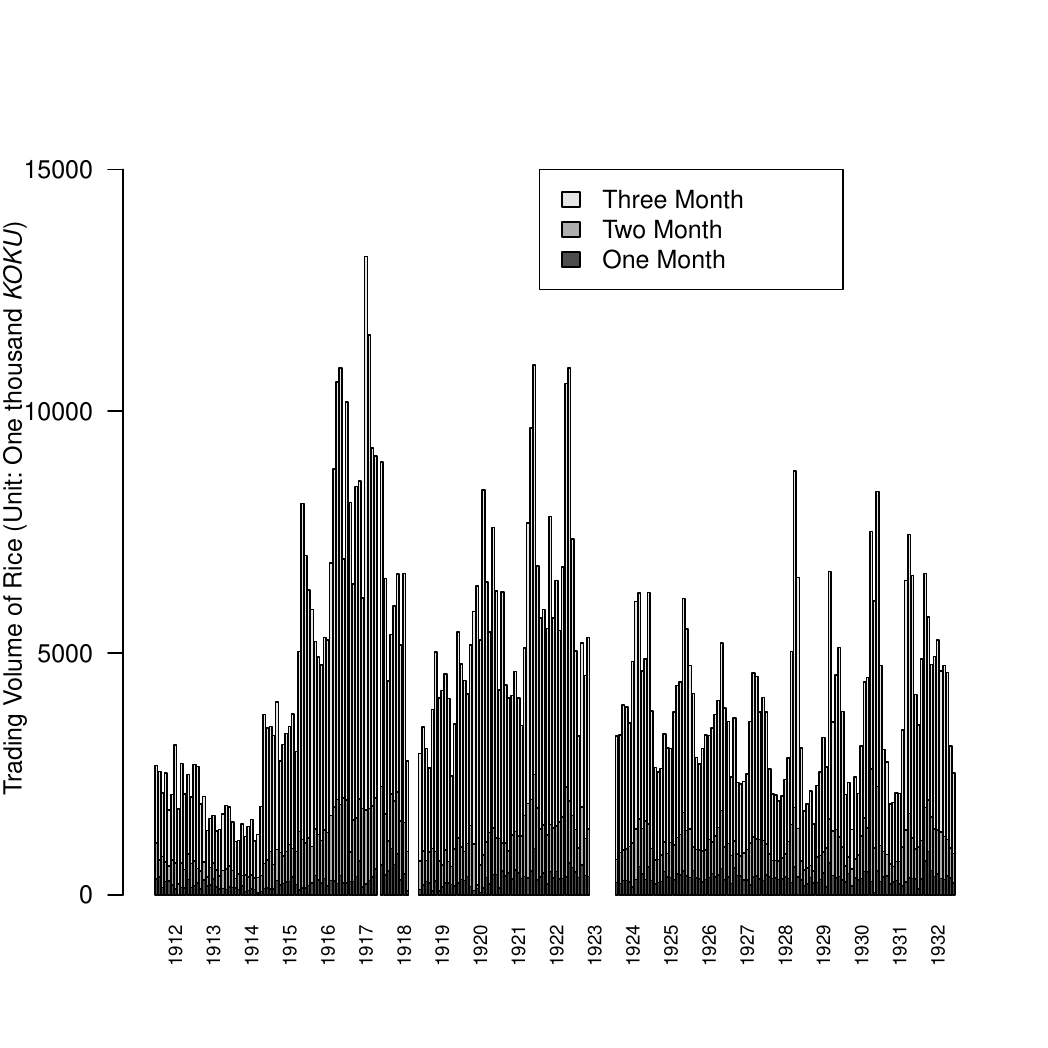}
\vspace*{3pt}
{
\begin{minipage}{350pt}
\footnotesize
\underline{Data Source}: \citet{maf1935tvp1}.
\end{minipage}}%
\end{center}
\end{figure}

\clearpage

\begin{figure}[bp]
 \caption{Trading Volume of Rice in the Osaka-Dojima Rice Exchange}\label{fpe_fig4}
 \begin{center}
 \includegraphics[scale=0.9]{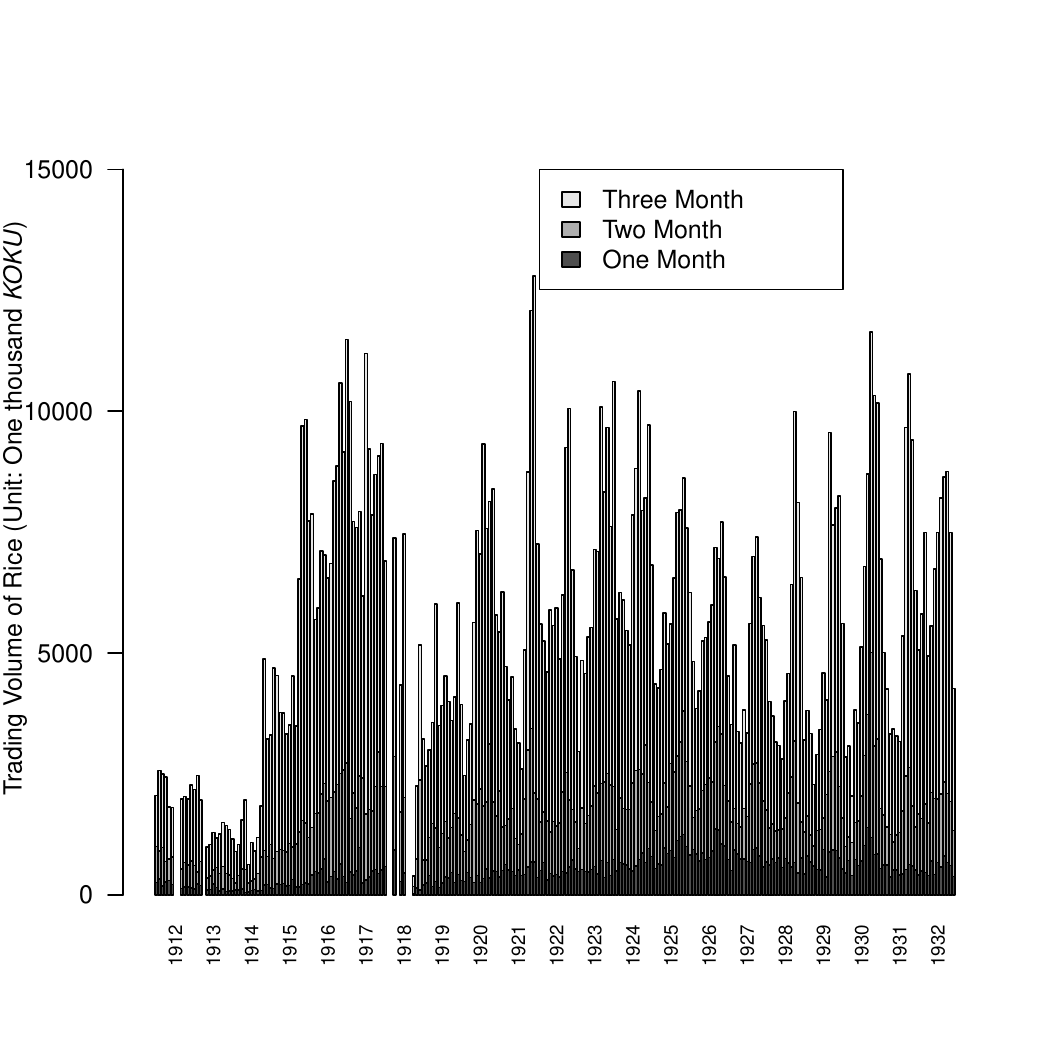}
\vspace*{3pt}
{
\begin{minipage}{350pt}
\footnotesize
\underline{Data Source}: \citet{maf1935tvp2}.
\end{minipage}}%
\end{center}
\end{figure}

\clearpage

\begin{figure}[bp]
 \caption{Delivery Volume in the Tokyo and Osaka-Dojima Rice Exchanges}\label{fpe_fig5}
 \begin{center}
 \includegraphics[scale=0.55]{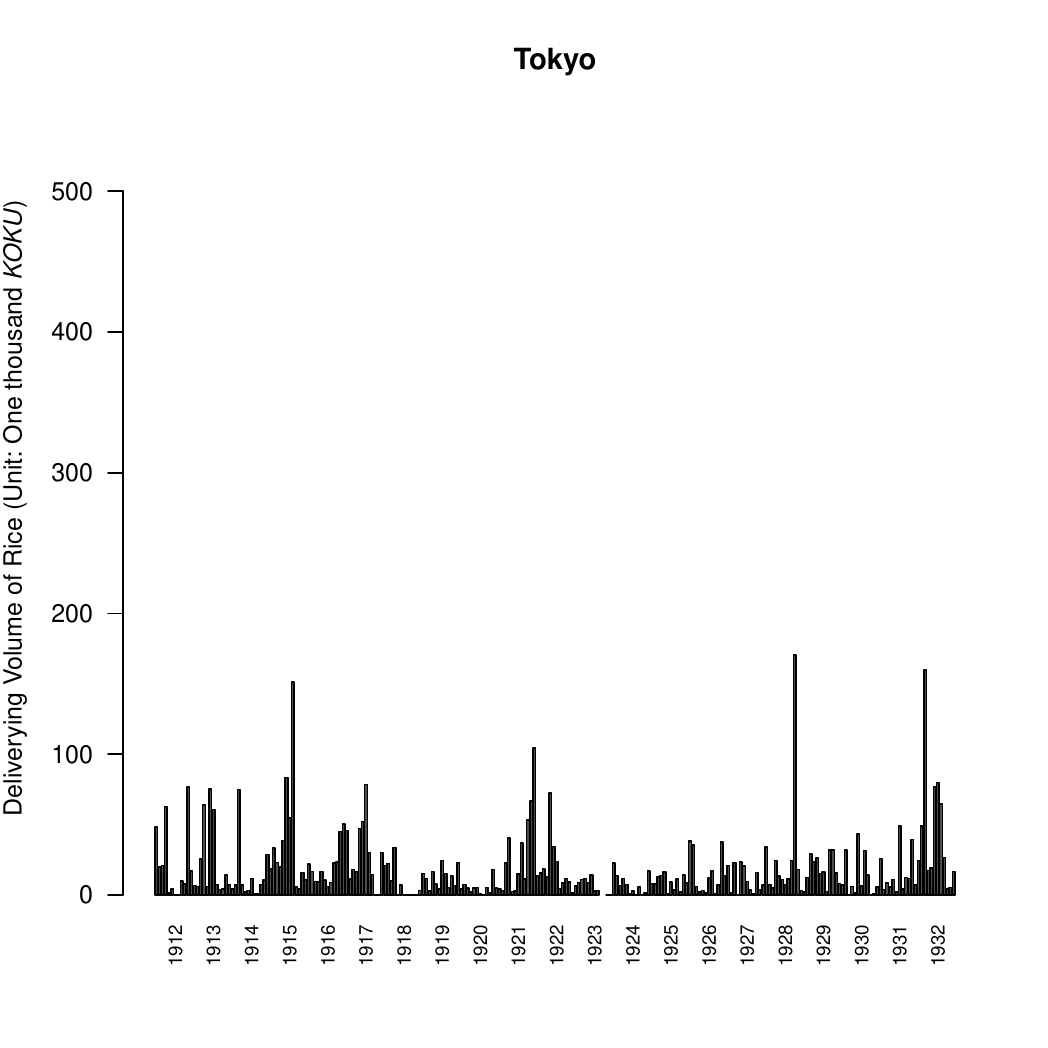}\\
 \includegraphics[scale=0.55]{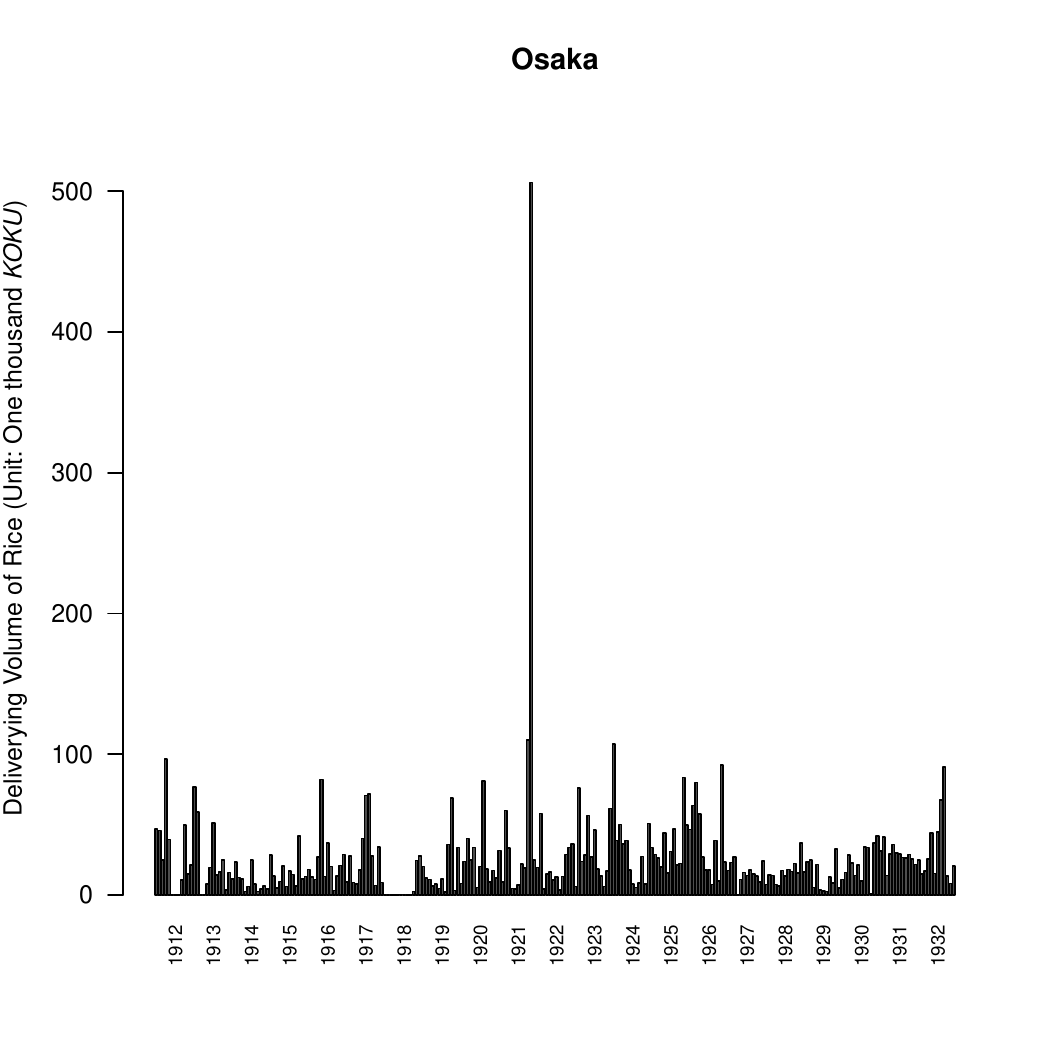}
 \vspace*{3pt}
{
\begin{minipage}{650pt}
\footnotesize
\underline{Data Sources}: \citet{maf1935tvp2,maf1935tvp3,maf1935tvp1}.
\end{minipage}}
\end{center}
\end{figure}

\clearpage

\begin{figure}[bp]
 \caption{Import, Export, and Consumption Volume of Rice in Japan (1880 to 1932)}\label{fpe_fig6}
 \begin{center}
 \includegraphics[scale=0.9]{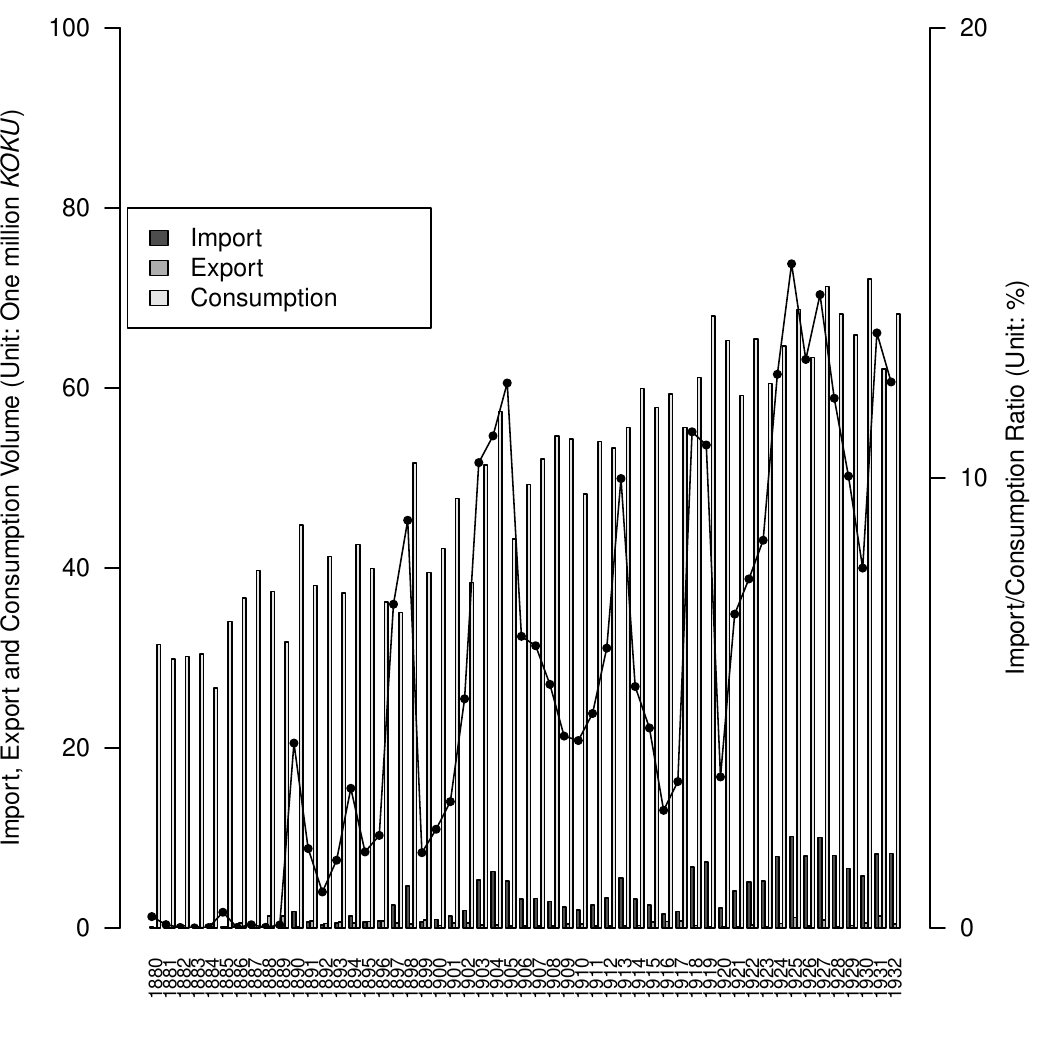}
\vspace*{3pt}
{
\begin{minipage}{350pt}
\footnotesize
\underline{Data Sources}: 
\begin{itemize}
 \item[(1)] \citet[pp.5,154,485,504,592]{toyo1935ftj}.
 \item[(2)] \citet[pp.108--109]{boj1966hys}.
\end{itemize}
\end{minipage}}%
\end{center}
\end{figure}

\clearpage

\begin{figure}[bp]
 \caption{Total Amounts of Rice Arrived in Tokyo (1912 to 1932)}\label{fpe_fig7}
 \begin{center}
 \includegraphics[scale=0.9]{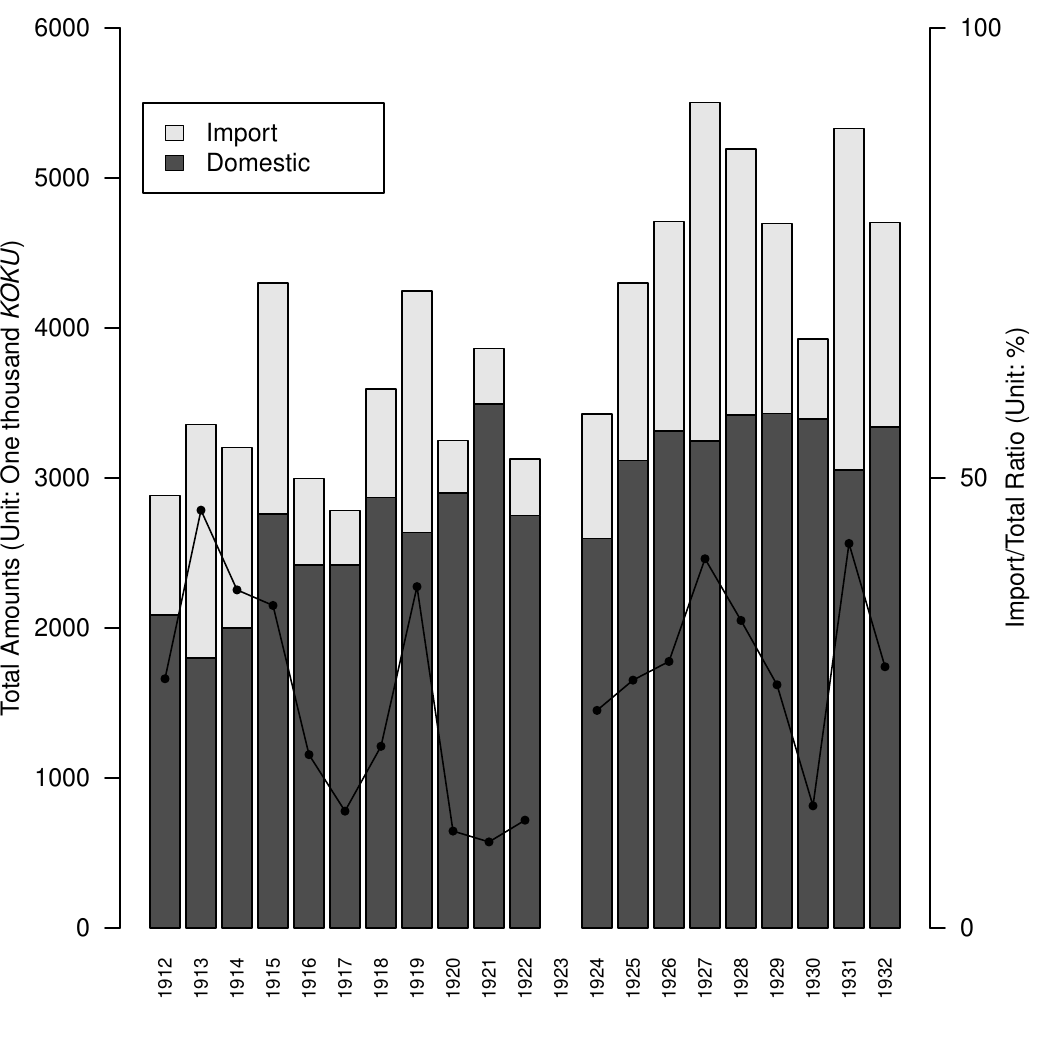}
\vspace*{3pt}
{
\begin{minipage}{350pt}
\footnotesize
\underline{Notes}: 
\begin{itemize}
 \item[(1)] The dataset is obtained from \citet[pp.273,281--282]{sasaki1937htr}.
 \item[(2)] There are no statistics for the amount of rice arrived in 1923 because the Great Kanto Earthquake occurred in Tokyo in September 1923.
\end{itemize}
\end{minipage}}%
\end{center}
\end{figure}

\clearpage

\begin{figure}[bp]
 \caption{Total Amounts of Rice Arrived in Osaka (1912 to 1932)}\label{fpe_fig8}
 \begin{center}
 \includegraphics[scale=0.9]{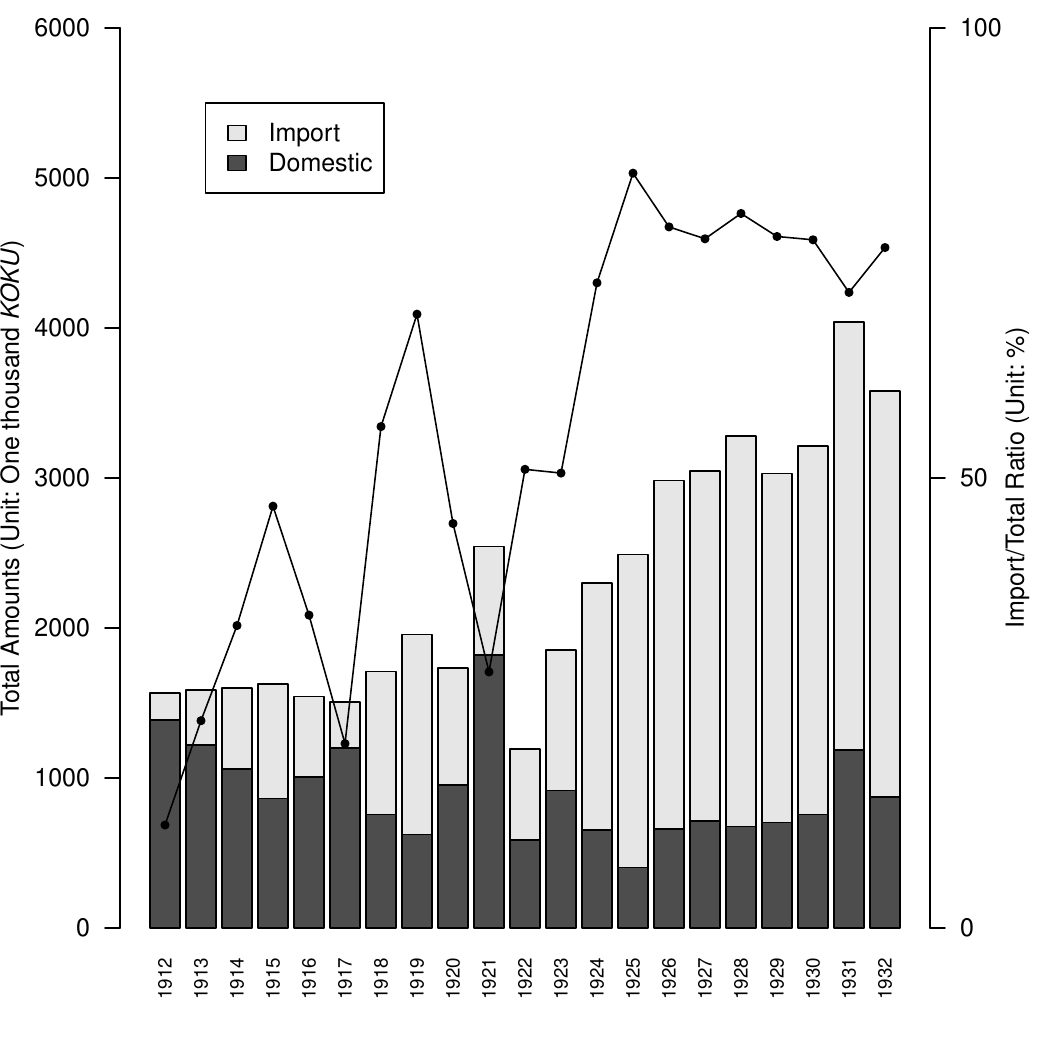}
\vspace*{3pt}
{
\begin{minipage}{350pt}
\footnotesize
\underline{Data Sources}: 
\begin{itemize}
 \item[(1)] \citet[ch.7, p.44]{osaka1919ocs}.
 \item[(2)] \citet[ch.7, pp.854--855]{osaka1925ocs}.
 \item[(3)] \citet[ch.5, pp.34--35]{osaka1933ocs}.
\end{itemize}
\end{minipage}}%
\end{center}
\end{figure}